\documentclass[11pt,a4paper]{article}
\usepackage{hyperref}
\usepackage{tikz-cd}
\usepackage[utf8]{inputenc}
\usepackage{graphicx}  \usepackage{mathptmx}
\usepackage{amsmath, amssymb}
\numberwithin{equation}{section}
\usepackage{gensymb}

\newcommand{\id}[1]{\ensuremath{\mathrm{id}}}

\newcommand{\half}{\mbox{\footnotesize $\frac{1}{2}$}}

\newcommand{\er}{\eqref}
 
\newcommand{\beq}{\begin{equation}}
\newcommand{\eeq}{\end{equation}} 
\newcommand{\bea}{\begin{eqnarray}}
\newcommand{\eea}{\end{eqnarray}} \newcommand{\nn}{\nonumber}

\newcommand{\ovl}{\overline}

\newcommand{\raw}{\rightarrow}

 \newcommand{\Raw}{\Rightarrow}

\newcommand{\la}{\langle} \newcommand{\ra}{\rangle}

\newcommand{\x}{\times}

\newcommand{\al}{\alpha} 
\newcommand{\gm}{\gamma} \newcommand{\Gm}{\Gamma}
\newcommand{\dl}{\delta} 
 \newcommand{\varep}{\varepsilon}

\newcommand{\rh}{\rho} \newcommand{\sg}{\sigma}
\newcommand{\Sg}{\Sigma}  
 \newcommand{\phv}{\varphi}
\newcommand{\ch}{\ch}  
\newcommand{\om}{\omega} \newcommand{\Om}{\Omega}

\newcommand{\inv}{^{-1}}

 \newcommand{\CF}{{\mathcal F}}

\newcommand{\CN}{{\mathcal N}} \newcommand{\CS}{{\mathcal S}}
\newcommand{\CO}{{\mathcal O}} 
 \newcommand{\CR}{{\mathcal R}}

\newcommand{\CZ}{{\mathcal Z}}
\newcommand{\N}{{\mathbb N}} \newcommand{\R}{{\mathbb R}}
\newcommand{\T}{{\mathbb T}} \newcommand{\Z}{{\mathbb Z}}
  \makeatletter
 
\makeatletter
\def\moverlay{\mathpalette\mov@rlay}
\def\mov@rlay#1#2{\leavevmode\vtop{%
   \baselineskip\z@skip \lineskiplimit-\maxdimen
   \ialign{\hfil$\m@th#1##$\hfil\cr#2\crcr}}}
\newcommand{\charfusion}[3][\mathord]{
    #1{\ifx#1\mathop\vphantom{#2}\fi
        \mathpalette\mov@rlay{#2\cr#3}
      }
    \ifx#1\mathop\expandafter\displaylimits\fi}
\makeatother

\newtheorem{definition}{Definition}[section]
\newtheorem{lemma}[definition]{Lemma}
\newtheorem{theorem}[definition]{Theorem}

\newcommand{\ML}{Martin-L\"{o}f}

\topmargin = - 1 cm \textheight = 23 cm \textwidth = 15 cm
\usepackage[symbol]{footmisc}
\renewcommand{\thefootnote}{\fnsymbol{footnote}}
\renewcommand{\Pr}{\mathrm{Prob}}
\newcommand{\CD}{\mathcal{D}}

\newcommand{\QED}{\null\nobreak\hfill\ensuremath{\square}}%
\newcommand{\BE}{Boltzmann equation}
\newcommand{\SZA}{\emph{Stosszahlansatz}}
\begin{document}
\pagenumbering{arabic} \setlength{\unitlength}{1cm}\cleardoublepage
\date\nodate
\begin{center}
\begin{Huge}
{\bf Irreversibility and  randomness}\end{Huge}
\bigskip\bigskip

\begin{Large}
 Nino Dekkers\footnote{Department of Mathematics and Computer Science, 
Technical University Eindhoven, The Netherlands, and Dutch Institute for Emergent Phenomena (DIEP), The Netherlands.
 \texttt{n.dekkers@tue.nl}.}
  and Klaas Landsman\footnote{ Radboud Center for Natural Philosophy (RCNP) and Department of Mathematics, Institute for Mathematics, Astrophysics, and Particle Physics, 
Radboud University,  Nijmegen, The Netherlands. \texttt{landsman@math.ru.nl}}
 \end{Large}\medskip
 \medskip

 \begin{abstract} 
\noindent
We  make the idea of ``molecular chaos'' precise through algorithmic randomness of microscopic trajectories. Apart from making this idea concrete, this has three advantages. First, it sharpens  ``typicality'' approaches to irreversible macroscopic evolution equations like the Boltzmann equation in giving a  criterion for \emph{individual} (as opposed to \emph{typical}) trajectories  to induce irreversible macroscopic behaviour. Second, algorithmic randomness theory comes with an effective ergodic theorem, which in our toy models yields the autonomy of macroscopic equations induced by microscopic data. Third, it relates macroscopic irreversibility of random trajectories to a lack of symmetry under time reversal of the  (algorithmic) criterion for randomness. This criterion is defined via an underlying probability measure $\mathbb{P}$ on the space of  microscopic trajectories. In  deterministic models these reduce to their initial conditions. Following most literature, including the
recent derivation of the  Boltzmann equation for long times, our $\mathbb{P}$ makes all particles i.i.d.\ at  $t=0$. 
After some qualitative comments on the Boltzmann equation in historical perspective, we realize our scenario in two toy models, viz.\ the (stochastic) Ehrenfest urn model and the (deterministic) Kac ring model. We finally discuss the relevance of Chaitin's incompleteness theorems, which (here) state the impossibility of explicitly displaying algorithmically random microscopic trajectories, despite their ubiquity. 
 \end{abstract}\end{center}
 \tableofcontents
 
\thispagestyle{empty}
\renewcommand{\thefootnote}{\arabic{footnote}}
\newpage \setcounter{footnote}{0}
\section{Introduction}
Despite approximately 150 years of effort and progress, starting with Boltzmann's great papers from 1872 and 1877, the problem of explaining the irreversibility of macroscopic  processes from an underlying reversible microscopic dynamics remains. We restrict ourselves to Boltzmann's probabilistic program, which is widely supported in both philosophy of physics and mathematical physics, yet requires considerable additional mathematical sophistication and conceptual clarification.\footnote{
See Boltzmann (1872, 1877) and the commentaries by Darrigol (2018), Uffink (2007, 2022),  Brown, Myrvold, and Uffink (2009), and Badino (2011).  For  general historical perspectives on irreversibility in the 19th century also see Brush (1974) and van Strien (2013). For later developments see Mackey (1992), Sklar (1993), and Zuchowski (2024).\label{fn1}}  
Our aim is to analyze the \emph{assumptions} behind the microscopic derivation of the Boltzmann equation (where we will not get beyond qualitative remarks) and some of its toy models (for which we will be complete), and introduce some perhaps unusual mathematical tools in this context.

On the mathematical front, in an impressive breakthrough Deng, Hani, and Ma (2024) recently advanced the work initiated by Grad (1958) and Lanford (1975, 1976) in giving a rigorous
derivation of the Boltzmann equation for  long times.\footnote{\label{mathrefs} Key textbooks between Lanford and Deng et al.\ include Spohn (1991), 
Cercignani, Illner, and Pulvirenti (2013), and Gallagher, Saint-Raymond, and Texier  (2014); see also the references in Deng, Hani, and Ma (2024), \S 1.2, and 
 Bodineau et al.\ (2026), \S 2.3. See footnote \ref{UV} for Valente (2014),  Uffink and Valente (2015), 
 and Ardourel (2017). }
The one-particle distribution function $f(t,\mathbf{z})$ which they prove to satisfy the Boltzmann equation under suitable assumptions is an \emph{average} (or expectation value) with respect to a certain probability measure on the space of microscopic degrees of freedom of a gas, which in this case is simply a Newtonian $N$-particle system, where $N\raw\infty$. This does not explain why the (initial) $N$-particle configuration of
  some \emph{given} gas induces a one-particle distribution function which in the limit $N\raw\infty$ satisfies the Boltzmann equation.\footnote{Compare interpretations of quantum mechanics that fail to explain why \emph{individual} measurements have outcomes.} That would require \emph{pointwise} results, initially  to the effect that the Boltzmann equation holds \emph{almost surely} (or ``typically'', as the neo-Boltzmannians have it).\footnote{What we identify as the neo-Boltzmannian (or ``typicality'') program is expounded by e.g.\  Lebowitz (1999), Lazarovici and Reichert (2015), and Bricmont (2020). The mathematical (physics) work from Grad (1958) onwards just referred to in effect  also follows this program. The core idea, which indeed may be attributed to Boltzmann (1872) but in a sense that could only later be recognized as being probabilistic, as in Boltzmann (1877), is that macroscopic quantities are coarse-grained functions of microstates and follow macroscopic equations for all ``typical'' microstates. In the thermodynamic limit, these and other ``typical'' properties hold almost surely with respect to some prior probability distribution on the microstates. Our work takes this idea one step further. See also the discussion after \er{MCH}.} 
   But even this would fail to identify exactly  \emph{which} microscopic configurations support the Boltzmann equation. This is the main problem we wish to address in this paper via the theory of algorithmic randomness. 
     
As a warm-up, let us review the strong law of large numbers for a fair coin toss, in which similar issues occur. Let $2^{\N}$ be the set of infinite binary sequences, equipped with the (fair) Bernoulli measure $f^{\N}$, see Appendix \ref{AR}, notably eqs.\ \er{deff} and \er{defqN}. For each $N\in\N_*$, define
  \begin{align} S_N: 2^{\N}\raw [0,1]; &&
  S_N(s):=
  \frac{1}{N}\sum_{n=0}^{N-1} s(n).\label{1.1}
  \end{align} 
  For later use, we note that in this specific case, $S_N$ is equivalent to the empirical measure 
    \begin{align} 
    \mu_N: 2^{\N}\raw \Pr(\{0,1\}); &&
  \mu_N(s):= \frac{1}{N}\sum_{n=0}^{N - 1} \dl_{s(n)}, \label{1.2}
  \end{align} 
  where $\Pr(X)$ is the set of  probability measures on $X$, here with $X=\{0,1\}$, and $\dl_a\in \Pr(X)$, defined for $a\in X$,  is the point measure defined on any (measurable) subset $U\subset X$ by
  \begin{align}
  \dl_a(U)=1 \hspace{20pt}  (a\in U); && \dl_a(U)=0 \hspace{20pt} (a\notin U).
  \end{align}For a finite set like $\{0,1\}$ it is easier to work with probability \emph{distributions} $p$
  defined on $\{0,1\}$  than with probability \emph{measures} defined on subsets of $\{0,1\}$;
 in that case, abusing notation,  $\dl_a$ in \er{1.2} is a function on $\{0,1\}$ namely 
  the Kronecker delta $\dl_a(b)=\dl_{ab}$. Since $a=\dl_a(1)$, we have 
\begin{equation}
S_N(s)=\mu_N(s)(1).
\end{equation}
The simplest result is $\la S_N\ra_{f^{\N}}\raw 1/2$, or, equivalently,  $\la \mu_N\ra_{f^{\N}}\raw f$,
 Since the empirical measure $\mu_N$ is a toy version of Boltzmann's one-particle distribution function,
this convergence of averages is conceptually similar to what has been proved for the Boltzmann equation (cf.\  \S\ref{OBE}). But convergence of averages contains no information about \emph{single} outcome sequences;  neither does convergence in probability. What about the strong law of large numbers? This law states that $f^{\N}$-almost surely,
\begin{align}  S_N(s)\raw 1/2; && \mu_N(s)\raw f .\label{slln}
\end{align}
 That is, one may remove a subset $\CN$ from $2^{\N}$ such that $f^{\N}(\CN)=0$ and \er{slln}  holds for all $s\in 2^{\N}\backslash \CN$ (i.e., $s\notin\CN$). But  $\CN$ is not given explicitly and one would like to know \emph{which}  $s\in 2^{\N}$ satisfy \er{slln}.
This question is addressed  by the theory of \emph{algorithmic randomness}, as follows:\footnote{Both bullets are due to P. \ML\ (1966). See also Calude (2002), Theorem 6.57. Our proofs are different.}
 \begin{itemize}
\item  Eqs.\ \er{slln} hold if $s\in 2^{\N}$ is $f^{\N}$\emph{-random}; a property defined in Appendix \ref{AR} as a special case of what we call \emph{$P$-randomness},\footnote{We use the letter $P$ for generic probability measures, and $f^\N$, and later $\mathbb{P}$ and variations thereof, for specific ones. } defined for so-called \emph{computable probability spaces} $(X,P)$.
\item The set $\CR$ of all $f^{\N}$-random sequences has $f^{\N}(\CR)=1$ (so one may take $\CN=2^\N\backslash\CR$).
\end{itemize}
The following line of proof of these results is instructive and  important, as it can be generalized. 
\begin{enumerate}
\item The theory of  \emph{large deviations} provides bounds like the following:\footnote{\label{LDrefs} First introductions to the theory of large deviations include Ellis (1995), Touchette (2009), and Landsman (2026b).  Ellis (1985), Dembo and Zeitouni (1998), and  Rassoul-Agha and Sepp\"{a}l\"{a}inen (2015) are  more advanced textbooks.
For a coin toss \er{LD1} follows from  Sanov's theorem or  Cram\'{e}r's theorem, which are equivalent in that case. } for any $\varep>0$, one has
\begin{align}
V_N(\varep):=\{s\in 2^{\N}:  |S_N(s)-1/2|>\varep\}; &&
f^{\N}(V_N(\varep))\leq e^{-c(\varep)N},\label{LD1}
\end{align}
for some $c(\varep)>0$ and sufficiently large $N$. 
\item The previous result yields 
\begin{equation}
\sum_{N=1}^{\infty} f^{\N}(V_N(\varep))<\infty, \label{1.6}
\end{equation} 
so that the Borel--Cantelli lemma applies:  $f^{\N}$-almost every $s\in 2^{\N}$ lies in only finitely many sets $V_N(\varep)$, so that for all $\varep>0$ one has $|S_N(s)-1/2| \leq \varep$ for all but finitely many $N$. This  implies the strong law of large numbers \er{slln}. See also footnote \ref{fn38} for further details. 
\item The theory of \emph{algorithmic randomness} (see Appendix \ref{AR}) gives the following lemma, which trivially follows from the specific definition of $P$-randomness we use (see Definition \ref{defCR}). It is therefore  a consequence of any other equivalent definition of $P$-randomness one may encounter (such as the one named after P.\ \ML)  and  hence is also valid for all computable probability spaces and ensuing notions of $P$-randomness we encounter in this paper:
\begin{lemma} \label{lemma1} If $(V_N)$ is a uniformly computable sequence of open subsets of $2^{\N}$ for which 
\begin{equation}
\sum_N f^{\N}(V_N)<\infty,
\end{equation} 
then each $f^{\N}$-random sequence $s\in 2^{\N}$ is contained only in finitely many $V_N$.
\end{lemma}
It trivially follows from \er{1.6} that for any fixed $\varep=1/m$, $m\in\N_*$,  the $V_N(\varep)$ form such a sequence (whose uniform computability, as defined in Appendix \ref{AR}, is easy to prove). Hence by the same argument that led to the strong law of large numbers \er{slln} we conclude that \emph{each} $f^{\N}$-random sequence $s\in 2^{\N}$ satisfies this law. The ``almost sure'' clause has disappeared!
\end{enumerate}
Adding dynamics and replacing the law of large numbers by some macroscopic equation (such as a toy \BE), this example will be a model for our reasoning throughout this paper.

Here is the plan for the remainder of this paper.
The full Boltzmann equation is discussed in \S\ref{OBE}. It is, at least for the moment, too complicated to be treated using algorithmic randomness, but we do try to provide clarity about other aspects relevant to the relationship between irreversibility and randomness in its context. Apart from being a goal by itself, this discussion also feeds the specific 
illustration of our approach in two well-known toy models of the Boltzmann equation, namely the (stochastic) Ehrenfest model and the (deterministic) Kac ring model.
The former will be discussed in \S\ref{EM}, whose results are almost entirely new,\footnote{Fun fact: these results were originally inspired by the work of A. \ML\ (1979), the brother of P. \ML.} confirming our picture that algorithmically random paths (as defined via a suitable probability measure on the space of paths that comes from the Markov property of the model) are irreversible. In fact, we will discuss both the original Ehrenfest model and a variant of it, which enables us to take certain limits more easily and which might even be physically preferable to the original one.\footnote{This is the modified Ehrenfest model introduced by Hauert, Nagler, and Schuster (2024).}
The latter (i.e., the Kac ring model) is the topic of \S\ref{KRM}, in which for completeness we rederive the pioneering results of Hiura and Sasa (2019).\footnote{Hiura \& Sasa (2019) introduced algorithmic randomness in kinetic theory. For other applications to physics see Zurek (1989), Svozil (1993, 2018), Calude (2004), Li \& Vit\'{a}nyi (2008), Landsman (2020, 2021, 2023), and Barrett, Chen, \& Lopez-Wild (2026). For philosophy see Porter (2012), Eagle (2021), and 
Barrett \& Chen (2025, 2026).
} In particular, we show how suitable randomness assumptions give rise to a toy Boltzmann equation.

In  \S\ref{EET} we give a new mathematical explanation of the autonomy of the toy Boltzmann equations of both models (in the sense that these  equations, though derived from microscopic data, in the limit only depend on these data via  the macroscopic quantities they induce) via \emph{effective ergodic theory}.\footnote{See especially Bienvenu et al.\ (2012). 
Footnote \ref{EETrefs} gives additional references. } As we will review, much as algorithmic randomness theory is able to point at individual sequences that satisfy the strong law of large numbers (namely the $f^\N$-random ones), it also sharpens  Birkhoff's pointwise ergodic theorem by identifying points $x$ in a computable probability space $(X,P)$ for which ergodic averages exists, namely, once again, the $P$-random ones. The ergodic theorem in question then precisely states that these averages are independent of $x$, and this is what yields the said autonomy in our toy models. This also turns out to be the right context for discussing our firm opinion that our randomness assumptions are \emph{sufficient} but not \emph{necessary} for the derivation of macroscopic equations, much as the property of $f^{\N}$-randomness is (much) stronger than necessary for obtaining the limiting frequency of $1/2$ for a fair coin toss. 

We end the main body of the paper with a Conclusion, in which we look back at our approach and analyze what is common among the three models we study, such as
the ubiquitous assumption that initial conditions be i.i.d. We identify the justification of this assumption as a major open problem, which, as we will explain,  has little or no connection with the past hypothesis. 
  We also discuss the role of idealizations (since all sharp results are based on limits $N\raw\infty$), which seem under control here, as well as the closely related place of Chaitin's incompleteness theorems about algorithmic randomness in our approach. Indeed, these theorems come in a finite-$N$ version and a (much less known) infinite-$N$ version, both of which provide  examples of undecidable statements in the sense of G\"{o}del's  first incompleteness theorem, and confirm that randomness is ``elusive''.
Finally, Appendix \ref{AR} summarizes those parts of the theory of algorithmic randomness we need, and Appendix \ref{AT} is a pr\'{e}cis on time reversal, again to the small extent we need this idea. 
 \section{Irreversibility of the  Boltzmann equation}\label{OBE}
 We start with a quick review of  Boltzmann's great papers from 1872 (in which he introduced the  transport equation named after him as well as his negentropic $H$-function) and 1877 (in which he began his  discretization and counting techniques.\footnote{Boltzmann (1877) also initiated the study of their asymptotics, which was one of the sources of the modern probabilistic theory of large deviations. See references in footnote \ref{LDrefs}, as well Ellis (1995)  in connection with Boltzmann.}  Although the second paper looks quite different from the first and indeed for Boltzmann himself was a fresh start,\footnote{See the references in footnote \ref{fn1}, as well as the summary below.}
 from a modern mathematical point of view the papers are  closely related: both are based on a coarse-graining technique now recognized as a passage to what we  call an \emph{empirical measure},\footnote{Our notation in  \er{deff} and \er{LN} is as follows.   For any  measurable space $X$ (equipped with a $\sg$-algebra $\Sg$ which we suppress), let $\Pr(X)$ be the
  space of probability measures on $X$. Any $x\in X$ defines a  \emph{point measure} $\dl_x\in\Pr(X)$ via $\dl_x(U)=1$ if $x\in U$ and $\dl_x(U)=0$ if $x\notin U$ (for any $U\in\Sg$). 
  In \er{deff} we have $X=\R^{2d}$, and in \er{LN} we have $X=A^N$.
  \label{fn15}} see also \er{1.2}.
  \begin{itemize}
\item Denoting  the space of probability measures on $\R^{2d}$ by $\Pr(\R^{2d})$,
Boltzmann (1872) associated a one-particle distribution function $f_N\in \Pr(\R^{2d})$ to a
 microscopic $N$-particle configurations
$(\mathbf{z}_1,  \ldots  \mathbf{z}_{N})\in\R^{2dN}$, where  $\mathbf{z}=(\mathbf{r}, \mathbf{v})$,\footnote{Boltzmann (1872) initially introduced
$f_N$ for homogeneous gases in which the spatial argument $\mathbf{r}$ is absent and the velocity $\mathbf{v}$ is replaced by the kinetic energy $x$, so that `the number of particles in unit volume whose kinetic energy at time $t$ lies between $x$ and $x+dx$ I will call $f(x,t)dx$' (p.\ 268 of the English translation). Later in the paper he introduces what we call $f_N(\mathbf{z})$ for inhomogeneous gases, with a similar interpretation. \label{f72}} We (re)write his construction as 
\begin{equation}
 f_N (\mathbf{z}_1, \ldots  ,\mathbf{z}_{N}):=
 \frac{1}{N}\sum_{n=1}^{N} \dl_{\mathbf{z}_n}. \label{deff}
\end{equation}
If we equip $\R^{2dN}$ with a probability measure $\mathbb{P}_N$, as we will, then, again from a modern point of view,  $f_N$ is a  random variable \emph{on} $(\R^{2dN}, \mathbb{P}_N)$ taking values \emph{in} $\Pr(\R^{2d})$. 
 In physics one prefers probability \emph{distributions} to \emph{measures}:
one would like to write some $\mu\in\Pr(\R^{2d})$ as $d\mu(\mathbf{z})= \rh(\mathbf{z})d\mathbf{z}$, where $\rh$ is a probability \emph{distribution} on $\R^{2d}$. This formally works for $\mu=\dl_{\mathbf{z}_n}$ if we take 
$\rh(\mathbf{z})=\dl(\mathbf{z}-\mathbf{z}_n)$, the Dirac delta-function, so that \er{deff} becomes
\begin{equation}
d f_N (\mathbf{z}_1, \ldots  ,\mathbf{z}_{N})(\mathbf{z})= \frac{1}{N}\sum_{n=1}^{N} \dl(\mathbf{z}-\mathbf{z}_n)d\mathbf{z}.\label{defff}
\end{equation}
In view of this, as long as we keep the official meaning \er{defff} in mind, we may also write
\begin{equation}
f_N (\mathbf{z};\mathbf{z}_1, \ldots  ,\mathbf{z}_{N})= \frac{1}{N}\sum_{n=1}^{N} \dl(\mathbf{z}-\mathbf{z}_n).\label{deff3}
\end{equation}
The point made after \er{deff} may then be rephrased as follows: $f_N$
 has two very different kinds of arguments, namely
$(\mathbf{z}_1, \ldots  ,\mathbf{z}_{N})$, making $f_N$ a random variable \emph{defined on}  $(\R^{2dN},\mathbb{P}_N)$, and $\mathbf{z}$,  making this random variable \emph{take values in} the space of probability distributions on $\R^{2d}$.
\item  
Boltzmann (1877) uses the same idea: in modern notation, let $A^N$ be the set of $N$-particle configurations $\sg:\{1, \ldots, N\}\raw A$, where $A$ is a finite set (whose elements Boltzmann took to be discrete energy levels),\footnote{Putting his energy unit $\varep=1$ for simplicity, Boltzmann (1877) has $A=\{0, 1, \ldots, p\}$. Writing $N$ for his $d$, his `complexions' $(k_1, \ldots, k_N)$ therefore correspond to our microstates $\sg\in A^N$.  His `distribution of states' (\emph{Zustandsverteilung}) $(N_0, \ldots, N_p)$  corresponds to our $Np_N$, so that $N_k=p_N(k)=\sum_{n=1}^{N} \dl_{\sg(n)k}$ is the number of particles with energy $k$. \label{B77}
} 
 and $N$ is once again the number of particles. One now has a discrete version $p_N$ of the  empirical measure $f_N$ in \er{deff}, which  coarse-grains $\sg\in A^N$ into
 \begin{align}
 p_N(\sg):= \frac{1}{N}\sum_{n=1}^{N} \dl_{\sg(n)}. \label{LN}
 \end{align}
Thus $p_N(\sg)$ is a probability distribution on $A$, and once again, if we equip $A^N$ with some probability measure $\mathbb{P}_N$, then $p_N$  is a  random variable \emph{on} $(A^N, \mathbb{P}_N)$
taking values \emph{in} $\Pr(A)$. Since $A$ is finite, the move from  measures to  distributions is trivial:\footnote{If $X$ is finite (and $\Sg$ is just the power set of $X$), then any probability \emph{measure} $\mathbb{P}$ on $X$ is equivalently given by a probability \emph{distribution} (or \emph{mass}) $P$, where $\mathbb{P}(U)=\sum_{x\in U} P(x)$ and conversely $P(x)=\mathbb{P}(\{x\})$.
  In that case, the probability distribution corresponding to the point measure $\dl_x$ is the Kronecker delta $\dl_x(y)=\dl_{xy}$. }
the analogue of \er{deff3}  is now simply
 given by $p_N(a;\sg)= \frac{1}{N}\sum_{n=1}^{N} \dl_{\sg(n)a}$, where $\dl_{ba}$ is the usual Kronecker delta. 
\end{itemize}
The above text was written with hindsight. Historically, apart from the jump from continuous to discrete variables 
there are two major differences between Boltzmann's  1872  and 1877 papers:
\begin{enumerate}
\item Although $f_N$ is a probability distribution on the one-particle phase space $\R^{2d}$  and Boltzmann (1872) stressed this probabilistic interpretation right from the start, he did not see $f_N$ as a random variable in its dependence on the $N$-particle coordinates $(\mathbf{z}_1, \ldots  ,\mathbf{z}_{N})$, since he did not assume a probability distribution on the latter.\footnote{Uffink (2007), p.\ 967, makes the following comment:  `It is clear that the conception of probability expounded here is thoroughly frequentist and that he takes `the laws of probability'  as empirical statements. Furthermore, probabilities can be fully expressed in mechanical terms: the probability distribution $f$ is nothing but the relative number of particles whose molecular states lie within certain limits. (\ldots)   Indeed, it seems to me that Boltzmann's emphasis on the crucial role of probability in this paper is only intended to convey that probability theory provides a particularly useful and appropriate language for discussing mechanical problems in gas theory. There is no indication in this paper yet that probability theory could play a role by furnishing assumptions of a non-mechanical nature, i.e., independent of the equations of motion.' Likewise, Darrigol (2018), p.\ xviii, takes Boltzmann's `probabilistic turn' to start only in 1876.} In contrast, Boltzmann (1877) explicitly assumes a uniform probability distribution on microstates, making $p_N$ a random variable. This was the real and deep introduction of probability into the problem of irreversibility. 

Furthermore, the distinction between micro- and macrostates, which Uffink (2007), p.\ 977, even calls `the decisive new element that allowed Boltzmann a complete reinterpretation of the notion and role of probability' is now explicit (whereas it was only implicit in 1872).
\item Nonetheless, in 1872 Boltzmann was closer (than in 1877)  to what one really wants, namely to describe the evolution of an individual gas from some initial state, as opposed to describing generic or typical behaviour of an ensemble (helpful as that may be, too).  Boltzmann's goal of introducing $f_N$ in 1872 was to study its time evolution (with an eye on entropy and irreversibility), which aspect, curiously,  was lacking for $p_N$ in 1877. This time  evolution was supposed to be governed by what we now call the \emph{Boltzmann equation} 
\begin{equation}
\partial_t f(t,\mathbf{z}) +\mathbf{v}\cdot \partial_{\mathbf{r}}f(t, \mathbf{z}) = C\left(f(t,\mathbf{z})\right), \label{Beq}
\end{equation} 
whose right-hand side is a quadratic  expression in $f$ incorporating two-body-collisions. As Boltzmann  showed through his $H$-theorem, eq.\ \er{Beq}  implies that $f$ evolves irreversibly. 
\end{enumerate}
Boltzmann derives \er{Beq} by considering how the one-particle distribution function \er{deff} changes under the effect of microscopic dynamics $\mathbf{z}_n \mapsto \mathbf{z}_n(t)$, which he took to be hard-sphere collisions. Thus (without such formulae) he considered the time-dependent one-particle distribution function
\begin{align}
f_N(t, \mathbf{z}_1, \ldots, \mathbf{z}_N) := \frac{1}{N}\sum_{n=1}^{N} \dl_{\mathbf{z}_n(t)} && \mbox{or} &&
f_N(t, \mathbf{z};\, \mathbf{z}_1, \ldots, \mathbf{z}_N) := \frac{1}{N}\sum_{n=1}^{N} \dl(\mathbf{z} - \mathbf{z}_n(t)),  \label{deff2}
\end{align}
and showed that it has autonomous  time evolution $f_N \mapsto f_N(t)$, i.e., not depending on the underlying microstate
$\mathbf{z}_1, \ldots, \mathbf{z}_N$ but only on $\mathbf{z}$. But he missed the point that \er{Beq} can only hold exactly in what we now called the Boltzmann--Grad limit (see below);  was also unclear about his assumptions, including even the precise nature of the one-particle distribution function (footnote \ref{f72}); and obviously lacked  mathematical rigour.\footnote{The need for mathematical rigour in deriving macroscopic laws from microscopic ones was formulated by Hilbert, who in the elaboration of his Sixth Problem stated that `Boltzmann's work on the principles of mechanics suggests the problem of developing mathematically the limiting processes, there merely indicated, which lead from the atomistic view to the laws of motion of continua.' (Hilbert, 1900/2000, p.\ 419). See also Ehrenfest and Ehrenfest (1911).} 
That said, taken in conjunction his papers from 1872 and 1877 were groundbreaking: most subsequent progress was based on a combination of Boltzmann's individual (i.e., single-system) spirit of 1872 with his probabilistic philosophy of 1877.  It is this combination that we also try to capture by adding algorithmic randomness to modern reasoning.

Until recently, the program of rigorously deriving the Boltzmann equation from hard-sphere dynamics was limited to short times
 (see footnote \ref{mathrefs}), where typically the average value of $f_N$ as defined in \er{deff2} was taken with respect to i.i.d.\ initial conditions.
  In a major step forward, Deng, Hani, and Ma (2024) extended these short-time derivations to arbitrary times for which solutions exist.   Their conceptual framework was the same as  Lanford's, who (as we already suggested)  combined  Boltzmann's ideas from 1872 and 1877 by interpreting his distribution function $f_N$ from 1872  as a random variable in its dependence on the microscopic variables $\{\mathbf{z}_n\}$; as we noted, this was how  Boltzmann saw his energy distribution function in 1877 (but not the one in 1872).
 
 For understanding our ``randomness first'' program, it is important to explain  what has been achieved  here mathematically,\footnote{See also Bodineau et al.\ (2026) for a summary. The  book-length survey by Villani (2002) remains worthwhile.} and hence to recognize what remains  unsatisfactory physically.\footnote{For simplicity  we ignore a complication: instead of taking the particle number  $N$ fixed and letting $N\raw\infty$,  Lanford (1975) and  Deng, Hani, and Ma (2024) use the  grand canonical ensemble and 
 take $N$ to be random variable whose expectation value is sent to infinity in the Boltzmann--Grad limit. This apparent complication in fact simplifies their proofs but could have been avoided (as they state themselves), and seems irrelevant for our purposes.}
  \begin{enumerate}
\item One assumes Newton's laws (for finite $N$) with hard-sphere dynamics, in which each particle moves freely until it encounters another particle, upon which they scatter elastically.
 Let $\varep$ be the diameter of the particles, with  non-overlapping domain $\CD_N\subset \R^{2dN}$ on which $|\mathbf{r}_n-\mathbf{r}_m|\geq \varep$ for all $1\leq n\neq m$. There is 
 a subset  $\CZ_N\subset \CD_N$ of  Lebesgue measure zero such that on $\CD_N\backslash\CZ_N$ the hard-sphere dynamics exists for all times; in particular,  collisions of more than two particles do not occur on  $\CD_N\backslash\CZ_N$.
 Since under \er{1.4} below, which is the basic assumption in all that follows,  initial data in $\CZ_N$ have zero probability even for finite $N$, 
to keep the notation simple we  assume that
$(\mathbf{z}_1, \ldots  ,\mathbf{z}_{N})\in \CD_N\backslash \CZ_N$ and henceforth ignore $\CZ_N$.
\item At $t=0$ the coordinates $\mathbf{z}_n$ of the $N$ particles are assumed independent and identically distributed (i.i.d.) on the non-overlapping domain $\CD_N$ via a probability density
\begin{equation}
p^{(N)}(\mathbf{z}_1, \ldots  ,\mathbf{z}_{N})=
C_N\prod_{n=1}^N p(\mathbf{z}_n)1_{ \CD_N}(\mathbf{z}_1, \ldots  ,\mathbf{z}_{N}) \label{1.4}
\end{equation}
on $\R^{2dN}$, 
where the density $p(\mathbf{z})$ on $\R^{2d}$  is some  one-particle distribution function,  assumed Lipschitz continuous in both variables with Gaussian decay in $\mathbf{v}$, and normalized such  that
\begin{equation}
\int_{\R^{2d}} d\mathbf{z}\, p(\mathbf{z})=1. \label{npz}
\end{equation}
Here
 $C_N>1$ is a normalization constant, which despite \er{npz}  is necessary because $\R^{2dN}\backslash \CD_N$ has positive Lebesgue measure. It will follow from point \ref{point4} below that the latter measure  vanishes as $N\raw\infty$, implying $C_N\raw 1$, so to simplify our expressions  we feel free to put $C_N=1$ in what follows. We likewise also omit the restriction $(\mathbf{z}_1, \ldots  ,\mathbf{z}_{N})\in  \CD_N$.
Thus the microstates $(\mathbf{z}_1, \ldots  ,\mathbf{z}_{N})$ become random variables with initial distribution \er{1.4}. 
We denote 
 the probability measure on $\R^{2dN}$ induced by the initial distribution \er{1.4} by $\mathbb{P}_N$.
\item \label{point4}  The function $f(t,\mathbf{z})$  that will eventually satisfy the Boltzmann equation is a limit $f_N\raw f$, where $f_N$ is given by \er{deff2}, whose initial value is the  function $p$ that occurs in \er{1.4}, i.e., 
\begin{equation}
f(0,\mathbf{z})= p(\mathbf{z}).\label{1.5}
\end{equation}
But what is meant by ``$f=\lim_{N\raw\infty} f_N$''? First, $\lim_{N\raw\infty}$ is in fact a double limit called the \emph{Boltzmann--Grad limit} (physically this is the \emph{dilute gas limit}), in which one  simultaneously takes $N\raw\infty$ and $\varep \raw 0$ at constant $\varep^{d-1}N$, where $d$ is the spatial dimension.

Second, since $f_N(t,\mathbf{z};\, \mathbf{z}_1, \ldots  ,\mathbf{z}_{N})$ as defined in \er{deff2} is a function of both $\mathbf{z}$ and  the random variable $(\mathbf{z}_1, \ldots  ,\mathbf{z}_{N})$, assumed distributed by \er{1.4}, whereas $f(t,\mathbf{z})$ is just a function of $\mathbf{z}$, one has to get rid of the dependence of $f_N$ on $(\mathbf{z}_1, \ldots,\mathbf{z}_{N})$. The most straightforward approach, which is used by Deng, Hani, and Ma (2024) and most of their predecessors,\footnote{Boltzmann himself excluded: in 1872 he did not average over microstates. Furthermore,
 Cercignani, Illner, and Pulvirenti (2013), \S4.6,   Villani (2013), based on  Sznitman (1991), and  Bodineau et al.\ (2023), Corollary 1.2.1, 
 prove convergence \emph{in probability} of $f_N(t,\cdot)$. As Bodineau et al.\ (2026),   explain on p.\ 16 of their review of the work of Deng, Hani, and Ma (2024),  the assumption of exchangeability all but eliminates the difference with convergence of averages.
 } is to take averages over $(\mathbf{z}_1, \ldots  ,\mathbf{z}_{N})$ with respect to the probability measure $\mathbb{P}_N$ on $\R^{2dN}$. That is, 
\begin{equation}
 f(t, \mathbf{z})=
\lim_{N\raw\infty} \langle f_N(t, \mathbf{z}) \rangle_{\mathbb{P}_N}. \label{limf1}
\end{equation}
But to be precise, one has to get rid of the distributional character of
$f_N(t,\mathbf{z};\, \mathbf{z}_1, \ldots  ,\mathbf{z}_{N})$ in $\mathbf{z}$ by smearing. Using \er{deff2},  the pairing of $f_N$, seen as a distribution in $\mathbf{z}$ keeping $(\mathbf{z}_1, \ldots  ,\mathbf{z}_{N})$ fixed, with a test function $h\in C_b(\R^{2d})$, is given by:\footnote{Since in our case $f_N$ is a distribution of order zero, i.e., a measure, one does not need \emph{smooth}  test functions.} 
\begin{equation}
\int_{\R^{2d}}  f_N(t, \mathbf{z}; \mathbf{z}_1, \ldots  ,\mathbf{z}_{N}) h(\mathbf{z}) d\mathbf{z} \equiv  \la  f_N (t, \cdot; \mathbf{z}_1, \ldots  ,\mathbf{z}_{N}),h\ra=N\inv \sum_{n=1}^{N}h(\mathbf{z}_n(t)),
\end{equation} where the middle notation would be the more official one, but in what follows we prefer the more informal left-hand side (which avoids  brackets). The limit $f_N \to f$ then reads
\begin{equation}
\left\langle \int_{\R^{2d}}  f_N(t, \mathbf{z}; \cdot) h(\mathbf{z})d\mathbf{z}\right\rangle_{\mathbb{P}_N}\raw 
\int_{\R^{2d}}  f(t, \mathbf{z}) h(\mathbf{z})d\mathbf{z},\label{MC}
\end{equation}
for each $h\in C_b(\R^{2d})$. This
weak convergence by smearing with $h$ is understood in \er{limf1}.
    \item The key result  of Deng, Hani, and Ma (2024), then,\footnote{What follows is is a (rigorous) simplification of the somewhat different way  Deng, Hani, and Ma (2024) state their result. First, they do not invoke our \er{deff2} but work with the lowest distribution function in the BBGKY hierarchy, which they denote by $f_1(t, \mathbf{z})$ and which  coincides with our $ \la f_N(\mathbf{z},t)\ra_{\mathbb{P}_N}$;
    see  Uffink (2007), argument leading to his eq.\ (164) on page 1037, or Bodineau et al.\ (2026), eqs.\ (15) - (17).
    This implies what we call the key result of Deng et al.\ (2024) above. Second, they  prove convergence $f_N\raw f$ in $L^1(\R^{2d})$, 
    which implies weak convergence in  $\mathrm{Prob}(\R^{2d})$. \label{average}}
      is that for all  times $t$ for which a solution of  the Boltzmann equation  \er{Beq} with initial value $p(\mathbf{z})$ exists, 
  the limit \er{limf1} exists in the sense \er{MC}, upon which the limit function  $f(t,\mathbf{z})$
satisfies the Boltzmann equation  \er{Beq} with initial value $p(\mathbf{z})$, which had  already defined $\mathbb{P}_N$ via \er{1.4}.
\end{enumerate}
A crucial part of their proof (as well as of all earlier approaches since Lanford) is \emph{propagation of chaos}: the independence assumption \er{1.4} at $t=0$ implies that at all (admissible) later times factorization of $p^{(N)}$ remains valid at least approximately. More precisely,  for any given $N$ and $k=1, \ldots, N$,  define $k$-particle functions $f_k^{(N)}(t,\mathbf{z}_1, \ldots,  \mathbf{z}_k)$ distributionally via the pairing
\begin{equation}
\int_{\R^{2dk}} f_k^{(N)}(t, \mathbf{z}_1, \ldots, \mathbf{z}_k)h(\mathbf{z}_1, \ldots, \mathbf{z}_k)d\mathbf{z}_1\cdots d\mathbf{z}_k:=\frac{1}{N^k}\left\la \sum^N_{n_1, \ldots, n_k=1}h(\mathbf{z}_{n_1}(t), \ldots, \mathbf{z}_{n_k}(t))
\right\ra_{\mathbb{P}_N},
\end{equation}
where  the sum is only over $k$-tuples $n_1, \ldots, n_k$ with distinct entries.
For example, eq.\  \er{1.4} gives
\begin{equation}
 f_k^{(N)}(0, \mathbf{z}_1, \ldots, \mathbf{z}_k)= p(\mathbf{z}_1)\cdots p(\mathbf{z}_k). \label{factor}
\end{equation}
Propagation of chaos then means that these $k$-particle functions factorize at all times, i.e.,\footnote{Eq.\ \er{MCH} follows from eq.\ (1.18) in Deng, Hani, and Ma (2024). }
\begin{equation}
\lim_{N\raw\infty} \left\|      f_k^{(N)}(t, \mathbf{z}_1, \ldots, \mathbf{z}_k)
-  \prod_{n=1}^k f(t,\mathbf{z}_n)\right\|_{L^1(\R^{2dk})}=0.\label{MCH}
\end{equation}

Summarizing the mathematical physics literature on the derivations and irreversibility of the \BE\ (see footnote \ref{mathrefs}), the latter arises from a \emph{combination} of four factors:\footnote{\label{UV} In our view, the confusion on the source(s) of the irreversibility of the \BE\ in  the literature since Lanford (1975, 1976) that has been diagnosed and analyzed by Valente (2014), Uffink \& Valente (2015), and Ardourel (2017)  has two sources: one is the conflation of irreversibility and the existence of an arrow of time, see the end of this section, and the other is the attempt to point at \emph{one} assumption in Lanford's theorems and their successors as \emph{the} source of irreversibility. The disagreements between Uffink \& Valente (2015) and Ardourel (2017) mainly concern problems with Lanford's approach that have meanwhile been straightened out and therefore will not be discussed here. } 
\begin{enumerate}
\item  \emph{coarse-graining}, introducing the one-particle distribution function $f(t,\mathbf{z})$ in the first place;
\item  \emph{special  initial conditions} expressing ``molecular chaos'',  making the particles i.i.d.\ at $t=0$;\footnote{Lanford (1976) and other authors cited in footnote \ref{mathrefs} 
 also considered some other possibilities, but in view of the success of Deng, Hani, and Ma (2024) we focus on the i.i.d.\ assumption. We will return to this in the Discussion.}
\item  \emph{taking suitable limits}, i.e., the Boltzmann--Grad limit $N\raw\infty$ and $\varep \raw 0$ at constant $\varep^{d-1}N$;
\item \emph{specific dynamics}, based on Newton's equations for a gas of hard spheres. 
\end{enumerate}
If one has all of these, then mathematically speaking the \BE\  comes out as  some complicated version of the law of large numbers.\footnote{See  Spohn (1991), \S 4.6, for a clear explanation of this conclusion, which goes back to Lanford (1975, 1976). }
The physical picture behind the mathematical derivation of the \BE\ may  be summarized as follows. 
The one-particle distribution function 
 $f_N(t,\cdot)$  develops and coarse-grains the initial microscopic conditions forward in time. Boltzmann's key observation was that two-body collisions contribute to the distribution function
  $f_N(t,\cdot)$ that solves what will be the \BE\  \emph{provided the particles in question are uncorrelated before they collide}.\footnote{Intuitively, correlations between two particles may come from  direction collisions of these particles in the past, or from earlier collisions with a third particle common to both, etc. This  is made rigorous using \emph{(collision) trees}, used by most authors  from Lanford (1975) to Deng, Hani, and Ma (2024). 
  See Bodineau et al.\ (2026) for an introduction.}
This \emph{proviso} then becomes a crucial \emph{assumption}, called the \SZA,
 which  Boltzmann (1877) however did not emphasize or even played down.\footnote{See Grad (1958), Uffink (2007, 2022),  Brown,  Myrvold, and Uffink (2009),  Uffink and Valente (2015), and Darrigol (2018)  for the relevant history and philosophy of the \SZA. Boltzmann (1872) had to assume the \SZA\ for all times, but this is no longer necessary in the modern derivations of the \BE.}

The initial distribution $\mathbb{P}_N$ in the modern derivations trivially enforces the \SZA\ at $t=0$, since $\mathbb{P}_N$ erases past correlations by the very choice of the  (i.i.d.) initial conditions. What is \emph{impossible} by assumption at $t=0$, namely the existence of correlations between particles about to collide, then still remains \emph{unlikely} at later times; in the Boltzmann--Grad limit  this possibility can even be ignored since it has zero probability.
This is what gives  `propagation of chaos': even after correlations have arisen, the probability that two \emph{already correlated} particles collide converges to zero,\footnote{This is very hard to prove rigorously, but it seems natural in the said limit, which makes the gas infinitely dilute.} and hence collisions that do take place with nonzero probability are  collisions between uncorrelated particles. Thus the \SZA\ and hence Boltzmann's equation continue to hold. 

We should now clearly distinguish between:
\begin{itemize}
\item Running the time-evolution of the  microscopic configuration $(\mathbf{r}_1(t), \mathbf{v}_1(t), \ldots  ,\mathbf{r}_{N}(t), \mathbf{v}_N(t))$ and hence of the induced one-particle distribution function $f_N(t, \mathbf{r}, \mathbf{v})$ as defined by \er{deff3} backward in time starting at some time $\tau>0$ by inverting all velocities and changing $t$ to $\tau-t$, that is, by considering $(\mathbf{r}_1(\tau-t), -\mathbf{v}_1(\tau-t), \ldots,\mathbf{r}_{N}(\tau-t), -\mathbf{v}_N)(\tau-t))$ for $t \in [0,\tau]$, with ensuing one-particle distribution function  $f_N(\tau-t, \mathbf{r}, -\mathbf{v})$. See also Appendix \ref{AT}.

In this case, the initial condition of the time-reversed show is the state at $t=\tau$ with velocities reversed, in which particles about to collide are likely to be correlated, so that the \SZA, and hence the \BE, fails. This is the sense in which this equation is \emph{irreversible}. 
\item Running the show backward in time from $t=0$ onwards, with the same (i.i.d.) initial conditions, by changing the sign of $t$ as well as reversing all velocities.

 In this case the \SZA\ is once again enforced by the initial condition and hence the \BE\ holds. This is the sense in which the entire scenario is \emph{invariant under time reversal}, so that \emph{the \BE\ provides no arrow of time.}
\end{itemize}
We therefore agree also on this point with Grad,\footnote{And also with Uffink \& Valente  (2015), Roberts (2022), and surely many others.
 \emph{Mutatis mutandis}, the discussion by Roberts (2022), Chapter 5,   partly based on Price (1996), is relevant here, e.g.:
`In our universe, an oscillating charge and a coordinated ring of oscillating charges are equally likely: if the first occurs, the second occurs as well, for example when the waves produced by the single oscillating charge are absorbed into the environment. More generally, the time reversal invariance of electromagnetism guarantees that if one solution is possible, then the time reversed solution is possible as well. One might like to add that an oscillating charge is more likely to occur `earlier than' a coordinated ring of charges in time. But, in which temporal direction are we looking when we say this? If we respond, ``towards (what we normally call) the future'', then we have just assumed what we were trying to prove, that there is a preferred direction of time for formulating such statements.'  (pp.\ 121--122). 
 The context
is the temporal asymmetry between the usual outgoing shells of electromagnetic waves produced by an oscillating charge, and incoming waves produced by a coordinated ring of oscillating charges, which are never observed. }
 who in our view described the situation  clearly:
 \begin{quote}
\begin{small}
Suppose we suddenly ``turn off''  \textsc{Liouville's} equation, reverse all velocities, and proceed again with the solution of \textsc{Liouville's} equation. We retrace our steps and find, for example, that [minus the entropy] $H_{\gamma}$ is increasing instead of decreasing. The reason is that, even though chaos is satisfied almost everywhere, the small exceptional set on which it is not satisfied has now become exactly the set on which the value of $f_2^{(N)}$ determines the course of $f$. (\ldots)
  
  As a final description of this situation we remark that \emph{there is no time's arrow}.
If we prepare a state we will observe that $H_{\gm}$, decreases; the same will be true if we follow the prepared state backwards in time. Or, if we observe an isolated system until it achieves a certain preassigned, unlikely state, we will observe that $H_{\gm}$ immediately begins to decrease; or if on observing this state, we reverse all velocities, we will again observe that $H_{\gm}$  decreases. No matter which direction in time we look, $H_{\gm}$ can only decrease. We can obtain an increase of $H_{\gm}$ if
we reverse the time only if we have specifically selected the given state not on its own merits but as the successor to another unlikely state.
(Grad, 1958, pp.\ 224,  228--229).
\end{small}
\end{quote}

The underlying mechanism of ``typicality'' may be  hidden in the use of averages in the definition \er{limf1} of the one-particle distribution function,  but such averages miss phenomena that happen in sets with very low (and, in the Boltzmann--Grad limit, zero) probability, which phenomena in this case consist of collisions between particles that are already correlated, which would compromise  the \BE. This is what ``typicality'' expresses, whose real thrust  is rather the exclusion of ``atypical'' phenomena!
This idea, still absent in Boltzmann (1872), goes back to Boltzmann (1877) at the latest, but it was left to later authors to (re)apply it to the \BE\ from 1872. First, in a physically  careful (but mathematically incomplete) discussion, Grad (1958), \S11, already made the crucial point in a quotation that preceded the previous one:
\begin{quote}
\begin{small}
It is possible to specify the exceptional set on which $f_2^{(N)}(\mathbf{z}_1, \mathbf{z}_2, t)$ does \emph{not} converge to $f(\mathbf{z}_1,t)f(\mathbf{z}_2, t)$ [as $N\raw\infty$] rather precisely and verify that this set can in fact be ignored.\footnote{What Grad means  is that the measure of these sets converges to 0 as $N\raw\infty$. His singular `set' may be confusing. } 
This exceptional set, on which chaos is not satisfied, has its own special interest. It consists, roughly speaking, of pairs of points $(\mathbf{z}_1,\mathbf{z}_2)$ that have recently collided. It is the existence of this exceptional set which provides the connection between the reversibility of \textsc{Liouville's} equation and the irreversibility of \textsc{Boltzmann's} equation. 
(Grad, 1958, p.\ 224) 
\end{small}
\end{quote}
Similarly, Spohn (1991), Chapter 9, explained this point as follows,\footnote{See also  Cercignani,  Illner, \& Pulvirenti (2013), \S4.7, for a detailed mathematical formulation of the same point.} switching his notation to ours and unfolding $\mathbf{z}=(\mathbf{r},\mathbf{v})$ at a crucial point in the quotation below. First, let $p(\mathbf{z})$ be a specific initial condition for the  \BE, with solution 
$f(t, \mathbf{z})$. Take some $\dl>0$ and let $\Gm_{\dl}\subset \R^{2dN}$ be the set of all $N$-particle configurations 
$(\mathbf{z}_1, \ldots  ,\mathbf{z}_{N})\in\R^{2dN}$ 
whose associated one-particle distribution function $f_N$ as defined by \er{deff3}, or more precisely the corresponding probability measure on $\R^{2d}$, is $\dl$-close to $f$  in $\Pr(\R^{2d})$, according to some metric.
\begin{quote}
\begin{small}
For some $(\mathbf{z}_1, \ldots  ,\mathbf{z}_{N}) \in \Gamma_{\dl}$ the [one-particle distribution function of the] curve $(\mathbf{z}_1(t), \ldots,\mathbf{z}_{N}(t))$ stays close to the solution of the Boltzmann equation $t\mapsto f(t,\mathbf{z})$ over a reasonable span of time. $(\mathbf{z}_1, \ldots  ,\mathbf{z}_{N})$  is then called a \emph{good} phase point. For other initial data, the \emph{bad} points in $\Gm_{\dl}$, however, the solution curve departs from the solution of the Boltzmann equation. The existence of bad phase points is a consequence of both the time-reversal invariance of Newton's equations of motion and the irreversible character of the Boltzmann equation: We choose a good configuration $(\mathbf{r}_1, \mathbf{v}_1\ldots  ,\mathbf{r}_{N}, \mathbf{v}_N)$ [whose associated  one-particle distribution function $f_N$ is] close to $p$  and evolve it to time $t > 0$. Then, by assumption the velocity reversed configuration $(\mathbf{r}_1(t), -\mathbf{v}_1(t)\ldots  ,\mathbf{r}_{N}(t), -\mathbf{v}_N(t))$ approximates well $f(t,\mathbf{r}, -\mathbf{v})$. In the future course of time the microscopic configuration retraces its own history and hence has to depart from the solution of the Boltzmann equation [with initial condition $f(t,\mathbf{r}, -\mathbf{v})$]. Thus $(\mathbf{r}_1(t), -\mathbf{v}_1(t)\ldots  ,\mathbf{r}_{N}(t), -\mathbf{v}_N(t))$ is a bad phase point for $f(t,\mathbf{r}, -\mathbf{v})$.

We infer that the good and bad phase points in $\Gm_{\dl}$  must be intermingled,
presumably in a complicated fashion. Besides the definition itself there is no
accesible criterion which would tell us whether a particular phase point $(\mathbf{z}_1, \ldots  ,\mathbf{z}_{N})\in\Gm_{\dl}$ 
is good or bad. We can offer only a \emph{measure-theoretical} characterization: ``Most''  phase points in 
$\Gm_{\dl}$ are good. The set of bad points has the less
weight the larger $N$ (\ldots). In other words: \emph{Typical}
phase points are good and, hence, follow the solution of the Boltzmann
equation.
We reemphasize that ``typical'' is understood here with respect to the initial
measure. (Spohn, 1991, pp.\ 151--152)
\end{small}
\end{quote}

This probabilistic scenario cannot, however,  be projected back onto  Boltzmann (1872), since he actually had the configuration of an individual gas in mind and so his distribution functions were defined by  \emph{counting} the number of particles with specific properties (regarding their positions and velocities). His idea of `correlation' was therefore the lack of factorization of the \emph{actual} number of particles in a certain two-body configuration as a product  two numbers defined by the corresponding one-body configurations.\footnote{See also  Brown,  Myrvold, and  Uffink (2009) and Uffink and Valente (2015).}
  
However, typicality does not say which \emph{individual} microstates  give rise to macroscopic evolutions (approximately) satisfying the Boltzmann equation, and hence to irreversible macroscopic behaviour. 
As illustrated in the Introduction for a fair coin toss, this deficiency can be solved in simple cases by appealing to algorithmic randomness. Since unfolding this idea for the \BE\ is far too hard mathematically for the time being, we now turn to some toy models which are accessible for this purpose. See also the Discussion for further comments.
\section{Irreversibility in the Ehrenfest model}\label{EM}
Predating their famous conceptual analysis of statistical mechanics (Ehrenfest and Ehrenfest, 1911), the oldest toy model for the \BE\ is what we now  call the \emph{Ehrenfest model} (Ehrenfest and Ehrenfest, 1907), which is subject of a large literature.\footnote{See e.g., quite differently,  Johnson and Kotz (1977), 
Baldovin, Caprini, and Vulpiani  (2019), and  Bricmont (2022).} We distinguish
 two versions, in each of which $N\in\N$ is  the total number of balls  that are distributed over  two different urns, labeled $U_0$ and $U_1$. Both are time-homogeneous Markov chains on finite state spaces.
\begin{itemize}
\item \emph{Microscopic version}:
this has state space $A_N=2^{N}$, so that a microstate $a\in A_N$ is a function 
\beq
a:\{0,\ldots, N - 1\}\raw \{0,1\},
\eeq which specifies the location of each ball:
$a_n\in\{0,1\}$  for each $n=0, \ldots, N - 1$ says that ball number $n$ is in urn $a_n$.  The transition probabilities are given by
\begin{align}
P_{ab}&=1/N  \mbox{ iff }b_n=a_n \mbox{ for all } n \mbox{ except one};\label{0.1}\\
P_{ab}&=0  \mbox{ otherwise}. \label{0.2} 
\end{align}  At each time step  a single ball is randomly and uniformly chosen from the set of $N$ balls, and is then moved to the other urn. This immediately gives \er{0.1} and \er{0.2}.
The stationary probability measure $\pi\in\Pr(A_N)$ is unique (by irreducibility) and is given by the flat distribution
 \begin{equation}
 \pi_a=2^{-N} \:\:\: \mbox{for each}\:\: a\in A_N.
 \label{pa2N} 
 \end{equation}
\item \emph{Macroscopic version}: here the state space is  $M_N=\{0,1,\ldots, N\}$, where $m\in M_N$ ($0\leq m\leq N$) is the number of balls in $U_1$ (so that $N-m$ is the number of balls in $U_0$). 
The transition probabilities come from the following picture, derived from the microscopic model: if  $U_1$ contains $m$ balls the probability that the jumping ball  is in $U_1$ is  $m/N$, in which case the  number of balls in $U_1$ after the jump equals $m-1$.The probability that the jumping ball is in $U_0$ equals $1-(m/N)$; after its jump to $U_1$ the  number of balls in $U_1$ is $m+1$.
This gives
\begin{align}
P_{m,m-1}:=\frac{m}{N} \:\:\: (m\geq 1); && P_{m,m+1}:=\frac{N-m}{N} \:\:\: (m<N); && P_{mn}=0\:\:\: (n\neq m\pm 1).
\label{0.5}
\end{align} 
The  stationary probability measure $\pi\in\Pr(M_N)$ is again unique, given by the binomial distribution
\begin{equation}
\pi_m=2^{-N}
\left(
\begin{array}{c}
N\\
m
\end{array}
\right). \label{binEF} 
\end{equation}
\end{itemize}
Though both versions may be defined separately, the macroscopic version can also be derived from the microscopic version. The connection is given by the coarse-graining map
\begin{align}
m: A_N\raw M_N; && m(a)=\sum_{n=0}^{N - 1}a_n. \label{mima1}
\end{align}
Denoting the microscopic Markov chain by $(X(t))_{t \in \N}$, governed by the transition probabilities \er{0.1} - \er{0.2}, and defining 
\begin{equation}
Y(t)=\sum_{n=0}^{N - 1}X_n(t) \label{mima2},
\end{equation}
the image of the microscopic process under \er{mima1}, the process  $(Y(t))_{t \in \N}$ is a Markov chain with transition probabilities \er{0.5}--note that an image of a Markov chain is not in general again Markov.

The one-particle distribution ``function'' of the Ehrenfest model, analogous to \er{deff}, is simply the coarse-graining map \er{mima1} divided by $N$, giving the proportion of balls in $U_1$. This is just a single number rather than a function, because a probability distribution on $\{0,1\}$ is determined by its value at 1. Adding time evolution, cf.\ \er{deff2}, we write
\begin{equation}
f_N(t) = \frac{1}{N}\sum_{n=0}^{N - 1}X_n(\lfloor Nt \rfloor), \label{mima3}
\end{equation}
for $t \in [0,\infty)$, where, following A. \ML\ (1979), we have rescaled the time variable, which is necessary to obtain a sensible limit as $N \to \infty$. Note that the random variable \er{mima3} depends on the \emph{path} of microstates (though we suppress this dependence in the notation for the time being, as is common in probability theory), unlike \er{deff2} which depends on a \emph{single} microstate. As before, we would like an autonomous evolution equation for $f_N(t)$, and the most straightforward way to achieve  this is to consider the average
\begin{equation}
\ovl{f}_N(t):=\la f_N(t)\ra_{\mathbb{P}_N}=\sum_{m=0}^N \frac{m}{N}\, \mathbb{P}_N\left(f_N(t)=\frac{m}{N}\right)\label{0.8}.
\end{equation}
Here $\mathbb{P}_N$ is the law of the microscopic process $(X(t))_{t \in \N}$ with $N$ particles (not scaled in time), which is determined by the transition probabilities \er{0.1} - \er{0.2} together with an initial distribution on $A_N$ at $t = 0$. Using \er{0.5}, we can derive the time evolution of $\ovl{f}_N$. If $Nt \in \N$, then
\begin{align}
\ovl{f}_N(t+1/N)&=\sum_{m=0}^N \frac{m}{N}\, \mathbb{P}_N\left(f_N(t+1/N)=\frac{m}{N}\right)
\nn \\ &= \sum_{\ell,m=0}^N \frac{m}{N}\, \mathbb{P}_N\left(f_N(t+1/N)=\frac{m}{N}\mid
f_N(t)=\frac{\ell}{N}\right) \mathbb{P}_N \left(f_N(t)=\frac{\ell}{N}\right)
\nn 
\\ &= \sum_{\ell,m=0}^N \frac{m}{N}\cdot P_{\ell m} \mathbb{P}_N \left(f_N(t)=\frac{\ell }{N}\right)
\nn \\ &=\sum_{\ell =0}^N \left(\frac{(\ell -1)}{N}\cdot\frac{\ell }{N}+\frac{(\ell +1)}{N}\cdot\frac{N-\ell }{N}\right) \mathbb{P}_N \left(f_N(t)=\frac{\ell }{N}\right)
\nn \\ 
&= \sum_{l = 0}^N \left(\frac{\ell}{N} + \frac{1}{N} - \frac{2\ell}{N^2}\right) \mathbb{P}_N \left(f_N(t)=\frac{\ell }{N}\right)  \nn \\
&=
\ovl{f}_N(t)+\frac{1}{N}(1-2\ovl{f}_N(t)).\label{0.9}
\end{align}
Solving this equation for $t$ such that $Nt \in \N$ and using $f_N(t) = f_N(\lfloor Nt \rfloor /N)$, we obtain 
\begin{equation}
\ovl{f}_N(t) - \frac{1}{2} = \left(\ovl{f}_N(0) - \frac{1}{2}\right) \left(1 - \frac{2}{N}\right)^{\lfloor Nt \rfloor}. \label{EhrenfestAverage}
\end{equation}
Provided $\lim_{N \raw \infty} \ovl{f}_N(0)$ exists, eq.\ \er{EhrenfestAverage} implies that $\ovl{f}(t) := \lim_{N \to \infty} \ovl{f}_N(t)$ also exists pointwise for any $t \in [0,\infty)$, and is given by
\begin{equation}
	\ovl{f}(t) - \frac{1}{2} = \left( \ovl{f}(0) - \frac{1}{2} \right)e^{-2t}, \label{EhrenfestAverageLimit}
\end{equation}
which exhibits an exponentially fast approach to the equilibrium value $\ovl{f}=1/2$ as $t\raw\infty$. One may also take $\lim_{N\raw\infty}$ before solving  \er{0.9} by rewriting it as
\begin{equation}
\frac{\ovl{f}_N(t+1/N)-\ovl{f}_N(t)}{1/N}=1-2\ovl{f}_N(t). \label{3.10}
\end{equation}
As  $N\raw\infty$, this yields, at least heuristically,   A.\ \ML's (1979)  toy Boltzmann equation
\begin{equation}
\frac{d\ovl{f}(t)}{dt}=1-2\ovl{f}(t). \label{0.10}
\end{equation}

Both the microscopic and the macroscopic Ehrenfest models are time symmetric in the specific way defined in Appendix \ref{AT}, since both satisfy \er{3.17b} and hence \er{3.17}; in both cases  the instantaneous time-reversal map $T$ is the identity. However, the fact that  our definition \er{3.17} of time-reversal invariance in time-homogeneous Markov chains forces the initial condition $\mu$ to be stationary (and hence maintains the system in equilibrium) may still leave room for irreversible behaviour if one picks a different initial condition. This does not happen in the microscopic Ehrenfest model, but in the macroscopic version, take some $\tau>0$ and for each $t\in [0,\tau]$ define  the time-reversed distribution function by
\begin{equation}
\ovl{f}_N^R(t):=\ovl{f}_N(\tau-t),
\end{equation}
 cf.\ \er{0.8} and \er{3.15}.
Using \er{EhrenfestAverageLimit}, the limit  
\beq
\ovl{f}^R(t) = \lim_{N \to \infty} \ovl{f}_N^R(t)
\eeq
 exists for any $t \in [0, \tau]$ and is given by 
\begin{equation}
	\ovl{f}^R(t) - \frac{1}{2} = \left(\ovl{f}^R(0) - \frac{1}{2}\right)e^{2t},
\end{equation}
which instead of \er{0.10} solves the toy anti-Boltzmann equation
\begin{equation}
\frac{d\ovl{f}^R(t)}{dt}=1+2\ovl{f}^R(t). \label{0.10b}
\end{equation}
In particular, this macroscopic theory fails to be time-reversal invariant in the sense that $\ovl{f}$ and $\ovl{f}^R$ satisfy different equations \er{0.10} and \er{0.10b}, respectively. As already noted in connection with the real \BE, this does not provide an ``arrow of time'': whatever properties $\ovl{f}$ has towards the future, $\ovl{f}^R$ has towards the past. The most telling such property is entropy increase: a natural (and successful)  choice here is the  Shannon entropy 
 \begin{equation}
S(t)= -\ovl{f}(t) \log \ovl{f}(t) - (1-\ovl{f}(t))\log (1-\ovl{f}(t)),
\end{equation}
where $\ovl{f}$ is seen a probability distribution $p$ on $\{0,1\}$ by identifying $\ovl{f}$ with $p(1)$;  this interpretation follows  from the definition of $f_N$ as the one-particle distribution function, i.e., $1/N$ times the number of balls in urn $U_1$, and the convergence $f_N\raw \ovl{f}$, which holds on average. Then
 \begin{equation}
\frac{dS}{dt}=(2\ovl{f}(t)-1)\log\left(\frac{\ovl{f}(t)}{1-\ovl{f}(t)}\right), 
\end{equation}
by the toy Boltzmann equation \er{0.10}. Hence $dS/dt\geq 0$ with $S$ reaching its maximum value at the equilibrium value $\ovl{f} = 1/2$, where $dS/dt=0$. All this is  analogous to the real Boltzmann equation.

Nonetheless, despite the fame of the Ehrenfest model and their success in modeling certain aspects of the real \BE, the underlying assumption that any at time  step \emph{exactly one ball jumps} is physically quite curious, for how are the other balls supposed to ``know'' that they must stay where they are? Randomness of the dynamics is no excuse, since this ``knowledge'' is supposed to exist for any sample path. In addition, the derivation of macroscopic evolution equations in the Ehrenfest model requires time scaling, analogous to the necessity of the Boltzmann--Grad limit in deriving the Boltzmann equation. As indicated at the end of \S 2, this forms a problem for our program of deriving macroscopic evolution equations from microscopic algorithmic randomness: there seems to be no version of the Ehrenfest model with $N = \infty$.\footnote{A possible solution to this problem is to restrict to a finite number of particles $N$ and to show that random paths of microstates satisfy \er{0.10} approximately. Since our main notion of algorithmic randomness, as given in Definition \ref{defCR}, is trivial for finite spaces, `randomness' should here be taken to mean \emph{Kolmogorov randomness} (see Appendix \ref{AR}). Results in this direction have already been obtained by Calude (2002), Chapter 5 for the law of large numbers for a finite number of fair coin tosses. Since macroscopic evolution equations are essentially consequences of the law of large numbers, we expect these results to generalize to the non-equilibrium setting.}

For these reasons we switch to a variant of the Ehrenfest model model introduced by Hauert, Nagler, and Schuster (2004), which we call the \emph{modified Ehrenfest model}. Here the state space remains $A_N=2^N$, again with the interpretation of balls jumping between two urns $U_0$ and $U_1$, but now each ball is independent in that for some fixed $\beta\in [0,1]$ it has its own transition probabilities, given by the $2\times 2$  matrix
\begin{equation}
(\pi_{ij})=\left(
\begin{array}{cc}
1-\beta  & \beta    \\
\beta  &   1-\beta
\end{array}
\right) \label{pvariant}
\end{equation}
and by definition the overall transition probability matrix between $a,b\in 2^N$ is given by
\begin{equation}
P^{(N)}_{ab}=\prod_{n=1}^{N} \pi_{a_nb_n}. \label{Pab}
\end{equation}
This defines a  Markov chain in the usual way upon supplying an initial distribution. The original Ehrenfest model is strictly speaking not a special case of this modified model, but the latter gives similar results upon setting $\beta=1/N$. This emerges most clearly in the macroscopic behaviour of the modified model. For example, if $U_1$ contains $m$ balls, then the probability of a single ball jumping from $U_1$ to $U_0$ is approximately $(m/N)/e$ and the probability of a single ball jumping from $U_0$ to $U_1$ is approximately $(1 - m/N)/e$. This differs from \er{0.5} by a factor $1/e$, stemming from the fact that transitions in which multiple balls jump are now also possible.

Unlike in the original Ehrenfest model,  in the modified model  rescaling of time is not necessary to derive a sensible macroscopic evolution equation in the limit $N \raw \infty$, since the transition probabilities \er{pvariant} do not scale with $N$, and we indeed do not rescale time so as to be able to take pointwise limits and refine these through algorithmic randomness. Consequently, the time evolution of the ``one-particle distribution''
\begin{equation}
	f_N(t) = \frac{1}{N} \sum_{n = 0}^{N - 1} X_n(t),
\end{equation}
where $t \in \N$, will be a difference equation rather than a differential equation. As before, we first derive the time evolution of the average 
\beq
\ovl{f}_N(t) = \la f_N(t) \ra_{\mathbb{P}_N},
\eeq
 where $\mathbb{P}_N$ is  the law of $(X(t))_{t \in \N}$ with an arbitrary initial condition. Since each $X_n(t)$ takes values in $\{0,1\}$, we may write
\begin{equation}
	\ovl{f}_N(t) = \frac{1}{N} \sum_{n = 0}^{N - 1} \mathbb{P}_N(X_n(t) = 1). \label{mem_average}
\end{equation}
Since  $(X_n(t))_{t \in \N}$ is a Markov chain with transition probabilities \er{pvariant} for each $n$, we have 
\begin{align}
	\mathbb{P}_N(X_n(t + 1) = 1) &= \mathbb{P}_N(X_n(t) = 0)\pi_{01} + \mathbb{P}_N(X_n(t) = 1)\pi_{11} \nn \\
	&= \mathbb{P}_N(X_n(t) = 1) + \beta(1 - 2\mathbb{P}_N(X_n(t) = 1)),
\end{align} 
which is a master equation.\footnote{For $\beta=1/2$ the master equation degenerates into $\mathbb{P}_N(X_n(t+1)=1)=1/2$, so that any initial distribution of $X_n(0)$ immediately collapses into equilibrium after a single time step.}
By linearity, this implies that \er{mem_average} satisfies the difference equation
\begin{equation}
	\ovl{f}_N(t + 1) - \ovl{f}_N(t) = \beta(1 - 2\ovl{f}_N(t)), \label{mem_evolution}
\end{equation}
which is solved by
\begin{equation}
	\ovl{f}_N(t) - \frac{1}{2} = \left(\ovl{f}_N(0) - \frac{1}{2}\right)(1 - 2\beta)^t. \label{mem_evolution_solution}
\end{equation}
If $\beta = 1/2$, this formula still holds for $t > 0$ if the convention $0^0 = 1$ is used. From now on we assume $\beta \in (0,1/2)$ to avoid pathological situations. With this assumption, \er{mem_evolution_solution} represents a monotonic and exponentially fast approach to equilibrium. Again, 
\beq
\ovl{f}(t) := \lim_{N \raw \infty} \ovl{f}_N(t)
\eeq
 exists if and only if $\ovl{f}(0) = \lim_{N \raw \infty} \ovl{f}_N(0)$ exists, and in that case $\ovl{f}$ satisfies 
\begin{equation}
	\ovl{f}(t + 1) - \ovl{f}(t) = \beta(1 - 2\ovl{f}(t)), \label{memtbe}
\end{equation}
which is the same as the toy Boltzmann equation \er{mem_evolution} but now without the subscript. 

As in the original Ehrenfest model and similarly to the real Boltzmann equation, the toy Boltzmann equation \er{memtbe} is an equation for the \emph{average} of the one-particle distribution function. But unlike the previous two cases, the modified Ehrenfest model allows us to formulate the dynamics directly with an infinite number of particles. The above strategy of starting with a Markov chain with state space $\{0,1\}$ and transition probabilities \er{pvariant}, and then taking a product of copies of this process also works to this end. The strong law of large numbers can then be formulated and proved, and may subsequently be refined by algorithmic randomness, as we now show. 

A stochastic process $(X(t))_{t \in \T}$ with state space $E$ may be represented using the product space $E^{\T}$ as underlying probability space, with a probability measure uniquely determined by the process (its \emph{Kolmogorov representation}).\footnote{This construction is completely standard in probability theory. See for example Dudley (1989), \S 8.2, or Klenke (2020), Chapter 14. Here $E^{\T}$ is equipped with the smallest $\sg$-algebra  $\mathcal{F}$ that makes each evaluation map $X_t:E^\mathbb{T}\raw E$, $x \mapsto x(t)$ measurable, given some $\sg$-algebra on the state space $E$ that is part of its definition as a measurable space.}  In particular, with $E = \{0,1\}$ and $\T = \N$ or an initial segment of $\N$, we let $(2^{\T}, \mathbb{P}_1)$ be the Kolmogorov representation of the two-state Markov chain with transition probabilities \er{pvariant} for some fixed $\beta\in (0,1/2)$ and initial distribution $p$ on $\{0,1\}$ defined by $p(1) = \alpha \in [0,1]$ (we keep the notation $\mathbb{T}$ in order to avoid confusion with another $\N$  below).
The modified Ehrenfest model with $N = \infty$ is then defined as the product of countably infinitely many copies of $(2^\mathbb{T},  \mathbb{P}_1)$, so that the ensuing probability space $(\Om,\mathbb{P})$ has
\begin{align}
 \Om=
 \prod_{n=0}^{\infty}2^\mathbb{T}=(2^\mathbb{T})^{\N}\cong (2^{\N})^\mathbb{T};
 && \mathbb{P} =\prod_{n=0}^{\infty}\mathbb{P}_1.
  \label{Omega}
\end{align}
The isomorphism means that we regard $2^{\N}$ as the state space of a stochastic process $(X(t))_{t \in \T}$ characterized by $\mathbb{P}$, with $(\Om, \mathbb{P})$ its Kolmogorov representation; in other words, whereas elements of $(2^\mathbb{T})^{\N}$ are by definition functions $\om:\N\raw 2^\mathbb{T}$, we see these as functions $\om:\mathbb{T}\raw 2^{\N}$. By construction, the process $(X(t))_{t \in \T}$ has the following properties:
\begin{itemize}
\item  At each fixed $n\in\N$, the $(X_n(t))_{t\in \mathbb{T}}$ form a two-state Markov chain in $\{0,1\}$ with transition probabilities \er{pvariant} and initial distribution $p$ such that $p(1) = \alpha$;
\item For each fixed $t\in \mathbb{T}$, the $(X_n(t))_{n \in \N}$ are i.i.d.
\end{itemize}
Because we have an infinite number of particles, the one-particle distribution function is  a limit 
\begin{align}
f(t,\om)=\lim_{N \to \infty} f_N(t, \om); && f_N(t,\om) = \frac{1}{N} \sum_{n = 0}^{N - 1} X_n(t,\om), \label{3.30}
\end{align}
where we have reinstated the dependence on the microscopic path $\omega$ in the notation, cf.\ \er{deff2},\footnote{Due to the construction \er{Omega}, the random variable $X_n(t)$ is simply the evaluation function $X_n(t,\om) = \om_n(t)$; the real content of the microscopic process lies in the probability measure $\mathbb{P}$.} and
the notation $f(t,\om)$ is reserved for those $\om$ for which the limit exists. This is not the case for every path $\om \in \Om$, but $f(t,\om)$ does exist for every $\mathbb{P}$-random path, as we will now show.

Using the theory of large deviations, we can prove that for $\mathbb{P}$-almost all $\om \in \Om$ this limit converges to a function $f(t)$ satisfying the toy Boltzmann equation
\begin{equation}
	f(t + 1) - f(t) = \beta(1 - 2f(t)). \label{tbe}
\end{equation}
More precisely, Cram\'{e}r's theorem, which we already used to derive \er{LD1}, in which $q$ is given by $q(0)=q(1)=1/2$, also works for possibly biased priors $q$ and gives almost identical results as the fair case: for any $m\in\N_*$, one has
\begin{align}
V_N(m):=\{s\in 2^{\N}:  |S_N(s)-q|>1/m\}; &&
q^{\N}(V_N(m))\leq e^{-d(m)N},\label{LD2}
\end{align}
for some $d(m)>0$ and sufficiently large $N$, using the same notation as in \S 1. Now define
\begin{equation}
	f(t) = \la f_N(t,\cdot) \ra_{\mathbb{P}}. \label{memdeff}
\end{equation}
By the same reasoning leading to \er{memtbe}, $f(t)$ satisfies \er{tbe}. By our choice of the initial condition of $\mathbb{P}_1$, $f(0) = \alpha$. Let $\T \subset \N$ be a finite initial segment. Applying Cram\'{e}r's theorem for biased priors to the sequence $(X_n(t))_{n \in \N}$, which is i.i.d.\ by construction, uniformly in $t \in \T$, we have 
\begin{align}
W_N(m):=\left\{\om\in (2^{\N})^\mathbb{T}: \| f_N(\cdot,\om)-f\|_{\infty}>1/m\right\}; 
&& \mathbb{P}(W_N(m))\leq e^{-d'(m)N},  \label{3.39}
\end{align}
for  $d'(m)>0$ and sufficiently large $N$, where  
\begin{equation}
\| f_N(\cdot,\om)-f\|_{\infty}:=\sup_{t\in \mathbb{T}} |f_N(t,\om)-f(t)|.
\end{equation} 
It then follows from the Borel--Cantelli lemma that $\mathbb{P}$-almost surely on $\Om$,
\begin{equation}
\lim_{N\raw\infty} f_N(\cdot,\omega) = f. \label{3.40}
\end{equation}
To strengthen the clause ``for $\mathbb{P}$-almost every $\om$'' in this result to our desirable clause ``for all  $\mathbb{P}$-random $\om$'' we may again use the   reasoning  based on
Lemma \ref{lemma1}, which gave us a similar strengthening of the strong law of large numbers  for a fair coin toss. As we noted, this lemma
holds for any computable probability space. Thus we obtain the main result in this section:\footnote{The computability requirements enabling Lemma \ref{lemma1} are easily checked. First, the topological space $(2^\mathbb{T})^{\N}$ becomes effective by numbering the cylinder sets. 
Second,  if $\alpha,\beta$ are computable, then so is $f(t)$ for all $t \in \mathbb{T}$, which  makes the family $(W_N(m))_N$ uniformly computable by saturating the inequality in its definition by computable opens.  }
\begin{theorem}\label{t3.1}
Let $\mathbb{P}\equiv \mathbb{P}_{(\al,\beta)}$ be the probability measure \er{Omega} on the space of paths $\Om$ of the modified Ehrenfest model with
initial condition $\alpha$ and transition probability $\beta$, both assumed to be computable reals. For each finite initial segment $\mathbb{T}\subset\N$ and each $\mathbb{P}$-random path $\om\in\Omega$, 
\begin{equation}
\lim_{N\raw\infty} f_N(\cdot,\om) = f,  \label{main}
\end{equation}
in norm, cf.\ \er{3.39}, where $f(t)$ is the solution of the toy \BE
\begin{equation}
 f(t+1) - f(t) = \beta(1-2f(t))\label{tBE}
\end{equation}
with initial condition $f(0) = \alpha$, for $t \in \T$.
\end{theorem}
In other words, $\mathbb{P}$-randomness (as defined in Appendix \ref{AR}) of a \emph{microscopic} path $\om$  is sufficient for the convergence of the corresponding \emph{macroscopic} truncated distribution function $f_N(\cdot,\om)$  to a solution of the toy \BE, where $\mathbb{P}$ is a probability measure on paths developing from i.i.d.\ initial conditions via the (Markov\-ian) stochastic dynamics \er{pvariant} - \er{Pab}.\smallskip

\emph{Proof}. This just summarizes our reasoning so far. Large deviations theory gives \er{LD2} applied to the sequence $(X_n(t))_{n \in \N}$ at fixed $t \in \T$ with probabililty distribution $q = f(t)$ on $\{0,1\}$, and hence \er{3.39}, since only finitely many times are involved. The appearance of $f$ is explained by \er{memdeff} and the argument leading to \er{memtbe} showed that $f$ satisfies \er{tBE}. The inequality in \er{3.39} enables Lemma \ref{lemma1},  which gives the implication
\begin{align}
	\omega\in (2^{\N})^\mathbb{T} \mbox{ is } \mathbb{P}\mbox{-random} &\Raw \om\in W_N(m) \mbox{ for finitely many }N.\label{endim}
\end{align} 
Since $m$ is arbitrary,\footnote{Since the finite subset $S\subset\N$ for which $\om\in W_N(m)$ iff $N\in S$ depends on $m$, let us be careful: the implication \er{endim} is $\forall_m\exists_{S\subset\N_f}\forall_{N\notin S} \forall_{t\in\T} |f_N(t,\om)-f(t)|\leq 1/m$, where $\T$ is finite. For all $m$ and $t$ this gives $\lim\sup_{N\raw\infty} f_N(t,\om)\leq f(t)+1/m$ and $\lim\inf_{N\raw\infty} f_N(t,\om)\geq f(t)-1/m$, so that 
$\lim_{N\raw\infty} f_N(t,\om)= f(t)$ for all $t\in\T$, which is \er{main}. \label{fn38}}
 this then implies \er{main}. \QED 
\medskip

What are the implications for time reversal? It would be natural to expect that $\mathbb{P}$-random\-ness of (microscopic) paths should not be invariant under time reversal, even in such a way that whereas the set of $\mathbb{P}$-random paths has probability one,  the set of time-reversed $\mathbb{P}$-random paths should have probability zero, which suggests why they are never observed. This indeed holds in our modified Ehrenfest model, but for the trivial reason that $\mathbb{P}$-randomness enforces the initial macroscopic condition $\alpha$. Writing $\T = \{0,1,2,\ldots,\tau\}$, the time reversal $\om^R$ of a path $\om \in \Om$ is 
\beq
\om^R(t) = \om(\tau - t),
\eeq
 cf.\ \er{3.15} with $T = \mathrm{id}$, and satisfies $f_N(t,\om^R) = f_N(\tau - t, \om)$. If $\om$ is $\mathbb{P}$-random, then 
\begin{equation}\label{reversed}
	\lim_{N \to \infty} f_N(t,\om^R) = f(\tau - t)
\end{equation} 
by Theorem \ref{t3.1}, with $f(t)$ the solution of the toy Boltzmann equation \er{tBE} with initial condition $f(0) = \alpha$. It follows that $\omega^R$ cannot be $\mathbb{P}$-random, since it has the wrong initial value $\lim_{N\to\infty} f_N(0, \omega^R) = f(\tau)$ (unless $\alpha = 1/2$, in which case $f$ is constant). Denoting the dependence of $\mathbb{P}$ on the initial macroscopic condition $\alpha$ explicitly, we can make the following less trivial assertion: if $\omega$ is $\mathbb{P}_{\alpha}$-random, then its time reversal $\omega^R$ is not $\mathbb{P}_{f(\tau)}$-random. With respect to this latter probability measure, the time-reversed macroscopic evolution \er{reversed} has the correct initial condition, but it does not satisfy \er{tBE} and hence $\omega^R$ cannot be $\mathbb{P}_{f(\tau)}$-random by Theorem \ref{t3.1}. Moreover, using \er{3.39} with $f(\tau)$ instead of $\alpha$ as initial condition, there is some $\varep > 0$ such that $\om^R\in W_N(\varep)$ for all sufficiently large $N$, so that such time-reversed paths, like any deviation from the expected macroscopic behaviour, are exponentially unlikely in $N$ as $N\raw\infty$.

To some extent this mirrors the role of  typicality in our discussion of the irreversibility of the \BE\ after \er{MCH}, especially as given in the quote by Spohn.\footnote{See also Theorem 5.1 in Hiura \& Sasa (2019), which is a similar result for the Kac ring model.}  In particular, in both cases the lack of time-reversal invariance of $\mathbb{P}$ is, crucially,  \emph{relative to the (i.i.d.) initial conditions.}

On the other hand, as in the original Ehrenfest model the result \er{memtbe} on convergence of averages does not depend on the initial condition, except for the mild demand of convergence at $t = 0$. Moreover, Theorem \ref{t3.1} even holds without the assumption that $(X_n(t))_{t \in \T}$ has the same distribution for each $n \in \N$, which is baked into the construction \er{Omega}.\footnote{But this time one cannot use the (standard version of) Cram\'{e}r's theorem anymore to prove this.} Making the distribution of $X_n(0)$ deterministic for each $n \in \N$, this implies that \emph{any} initial microstate may lead to irreversible behaviour, which is a big difference with deterministic models. This raises the question what the role of the \SZA\ in the modified Ehrenfest model could be, either in Boltzmann's original form in or the modern probabilistic form (cf.\ \S\ref{OBE}). Let us discuss various perspectives on this.
\begin{enumerate}
\item While every initial microstate $s \in 2^{\N}$ may lead to irreversible behaviour, it does not follow that each path starting from $s$ behaves irreversibly, as obvious counterexamples show;  instead, the analogue of Boltzmann's original \SZA\ in the modified Ehrenfest model exists on the level of paths. Recalling the notation \er{3.30}, 
the \SZA\ states that \emph{the proportion of balls in $U_1$ which jump to $U_0$ in the next time step is $\beta f(t,\om)$, and similarly equals $\beta(1 - f(t,\om))$ for balls jumping from $U_0$ to $U_1$.} From this assumption, one  obtains
	\begin{align}
		f(t + 1,\omega) - f(t,\omega) &= \beta(1 - f(t,\omega)) - \beta f(t,\omega) \nn \\
		&= \beta(1 - 2f(t,\omega)),
	\end{align}
	which is our previously-derived toy Boltzmann equation.
	The \SZA\ of the modified Ehrenfest model can actually be rigorously formulated and proved to hold for all $\mathbb{P}$-random $\om \in \Om$, once again not necessarily assuming that the balls have identically distributed Markov chains. This rigorous formulation for balls jumping from $U_1$ to $U_0$ is 
	\begin{equation}
		\lim_{N \to \infty} \frac{1}{N} \sum_{n = 0}^{N - 1} X_n(t, \omega) (1 - X_n(t + 1, \omega)) = \beta\cdot\lim_{N \to \infty} \frac{1}{N} \sum_{n = 0}^{N - 1} X_n(t, \omega),
	\end{equation}
	which can be rewritten as
	\begin{equation}
		\lim_{N \to \infty} \frac{1}{N} \sum_{n = 0}^{N - 1} X_n(t,\omega)(Y_n(t,\omega) - \beta) = 0, \label{szalln}
	\end{equation}
	where $Y_n(t,\cdot)$ is the random variable with value $Y_n(t,\omega) = 1$ if the $n$'th ball jumps during the time step $t \raw t + 1$, and $Y_n(t,\omega) = 0$ otherwise. Clearly, $X_n(t,\cdot)$ and $Y_n(t,\cdot)$ are independent and $\la Y_n(t,\cdot) \ra_{\mathbb{P}} = \beta$. Hence one recognizes the structure of the law of large numbers in \er{szalln}. Indeed, just as in Theorem \ref{t3.1}
	this equality can be proved  for all $\mathbb{P}$-random $\om \in \Om$. 	
	\item While the assumption of i.i.d.\ initial conditions can be relaxed as indicated, independence does remain a crucial assumption. Even in the above-mentioned scenario with a fixed initial microstate, in which each $X_n(0)$ is deterministic, independence is satisfied (since deterministic random variables are trivially independent). In fact, under the assumption of exchangeability at $t = 0$, which is a natural assumption in view of our ignorance of microscopic conditions, i.i.d.\ initial conditions are \emph{necessary} for laws of large numbers such as \er{main}. 

In a more general setting, let $(X_n)_{n \in \N}$ be an exchangeable sequence of random variables taking values in $\{0,1\}$ with common mean $\mu$. In particular, exchangeability  implies that the random variables are identically distributed. Setting $S_N = N^{-1} \sum_{n = 0}^{N - 1} X_n$, the weak law of large numbers
	\begin{equation}
		\lim_{N\raw\infty} \mathbb{P}(\lvert S_N - \mu \rvert \geq \varep) = 0
	\end{equation}
	is then equivalent to $\lim_{N\raw\infty}\mathrm{Var}(S_N) = 0$; one implication makes use Chebyshev's inequality, the other follows from bounding $\mathrm{Var}(S_N)$ by $\mathbb{P}(\lvert S_N - \mu \rvert \geq \varep)$. Using exchangeability, 
	\begin{equation}
		\mathrm{Var}(S_N) = \frac{\sg^2}{N} + \left(1 - \frac{1}{N}\right)\rh,
	\end{equation}
	where $\sg^2 = \mathrm{Var}(X_1)$ and $\rh = \mathrm{Cov}(X_1,X_2)$. Hence $\lim_{N \raw \infty} \mathrm{Var}(S_N) = 0$ iff $\rho = 0$, which for random variables taking values in $\{0,1\}$ is equivalent to pairwise independence. Using de Finetti's theorem,\footnote{See Cifarelli and Muliere (2002) for an  introduction to de Finetti's theorem. See also Klenke (2020), \S\S 12.3, 13.4.}
	  this can be upgraded to mutual independence, from which we conclude that $(X_n)_{n \in \N}$ are i.i.d. To see this, use de Finetti's theorem to write
	\begin{equation}
		\mathbb{P}(X_0 = x_0, \ldots, X_{N - 1} = x_{N - 1}) = \int_0^1 \prod_{n = 0}^{N - 1} \theta^{x_i}(1 - \theta)^{1 - x_i}\, d\nu(\theta), \label{definetti}
	\end{equation}
	 for some probability measure $\nu$ on $[0,1]$. Using this, as well as  pairwise independence $\mathbb{P}(X_0 = x_0, X_1 = x_1) = \mathbb{P}(X_0 = x_0)\mathbb{P}(X_1 = x_1)$, we obtain
	\begin{equation}
		\int_0^1 \theta^2\, d\nu(\theta) = \left(\int_0^1 \theta\, d\nu(\theta)\right)^2.
	\end{equation}
	By the Cauchy-Schwarz inequality, this can only be the case if $\nu$ is deterministic, which implies that \er{definetti} factorizes, i.e., the random variables $(X_n)_{n \in \N}$ are mutually independent. Note that this result relies crucially on the equivalence between pairwise independence and vanishing covariance, which only holds for random variables taking values in $\{0,1\}$ (we will discuss the case of more general state spaces in \S\ref{EET}).

	Applying this reasoning to the modified Ehrenfest model, we see that \er{main} (which has the weak law of large numbers as a consequence) implies that $(X_n(0))_{n \in \N}$ are i.i.d., under the assumption of exchangeability. Note that this has nothing to do with the dynamics, since we only need to look at the initial time $t = 0$.	\end{enumerate}
\section{Irreversibility in the Kac ring model}\label{KRM}
 A similar scenario may be realized  in the deterministic Kac ring model.\footnote{The original reference for the Kac ring model is Kac (1959). Useful   literature includes 
Maes, Neto\u{c}n\'{y}, and Shergelashvili (2007),    De Bi\`{e}vre and  Parris (2017), and Jebeile (2020). Our take is quite different. } Since this model has already been studied from a similar point of view in the pioneering paper by Hiura and Sasa (2019), we will be brief. We rederive their main result and add some relevant insights. 
We again start with a microscopic description. For finite $N$, the microscopic state space of the Kac ring  is 
\begin{equation}
A_N:= 2^{2N+1}\x 2^{2N+1}, \label{KRstate}
\end{equation}
whose elements we write as pairs $(x,y)$ consisting of two kinds of microscopic degrees of freedom. The component $x_n$ represents the colour of a ball at the $n$'th site, which is white if $x_n = 1$ and black if $x_n = 0$. For the other kind of microscopic variables, we say that site $n$ is \emph{marked} if $y_n = 1$ and \emph{unmarked} if $y_n = 0$. The microscopic dynamics is defined by the following map:
\begin{align}
	\varphi : A_N \raw A_N; && \varphi(x,y)_n = (x_{n - 1}+ y_{n - 1}(1 - 2x_{n - 1}), y_n),\label{kacevolution}
\end{align}
whose $t$-fold iteration we denote by $\varphi_t$. The subscripts here are always taken modulo $N$, in effect enforcing periodic boundary conditions, whence the name `ring'. The interpretation of \er{kacevolution} is as follows. At each time step every ball moves one site  in a fixed direction of the ring. If it leaves a marked site its colour changes; otherwise its colour remains unchanged. The set of marked sites does not change. The dynamics of the Kac ring is reversible, cf.\ \er{defT}, with instantaneous time-reversal map $T:A_N\raw A_N$ given by
\begin{equation}
	T(x,y)_n = (x_{-n}, y_{-n-1}). \label{TKac}
\end{equation}
The Kac ring model may also conveniently be described using the coordinates $(\eta,\varep) $ defined by
\begin{align}
(\eta,\varep) \in \{-1,1\}^{2N + 1} \x \{-1,1\}^{2N + 1}; &&
	\eta_n = 2x_n - 1 ; &&
	\varep_n = -2y_n + 1.
\end{align}
In this representation, $\eta$ is often interpreted as a configuration of spins and $\varep$ as a configuration of scatterers. Using these variables, the dynamics becomes much simpler:
\begin{align}
	\varphi(\eta,\varep)_n = (\eta_{n - 1}\varep_{n-  1}, \varep_n); && 	\varphi_t(\eta,\varep)_n = (\eta_{n - t}\varep_{n - t}\dotsb\varep_{n - 1}, \varep_n). \label{alternative}
\end{align}
Because the Kac ring has two kinds of microscopic degrees of freedom, there are two 
one-particle distribution functions acting as macroscopic variables, one for balls and the other for markers: 
\begin{align}
f_N(x, y):=\frac{1}{2N+1}\sum_{n=-N}^N x_n; && s_N(x, y)=\frac{1}{2N+1}\sum_{n=-N}^N y_n. \label{4.4}
\end{align}
The first gives the proportion of white balls, the second the proportion of marked sites. Note that the latter is non-dynamical, in the sense that it is unchanged under the dynamics. Writing $\varphi_t(x,y)_n = (x_n(t), y_n(t))$, with of course $y_n(t) = y_n$, the time evolution of $f_N$ reads
\begin{equation}
	f_N(t, x, y) := f_N(\varphi_t(x,y)) = \frac{1}{2N+1}\sum_{n=-N}^N x_n(t). \label{4.4evolution}
\end{equation}
Scaling in time, though an option, is not necessary to obtain limits of these macroscopic quantities as $N \raw \infty$. As in all previous cases, we introduce probability by assuming an i.i.d.\ initial distribution over microstates $(x,y)$, in which pairs $(x_n,y_n)$  are i.i.d.\ as $n$ runs through the $2N + 1$ indices, with $x_n$ and $y_n$ also independent of each other. Thus we equip $A_N$ with the product Bernoulli measure 
\begin{equation}
\mathbb{P}_N=\al^{2N+1}\times \beta^{2N+1},\label{PNalbeta}
\end{equation}
where we label probability measures on $\{0,1\}$ by their means. As in the Ehrenfest model, we will assume $\beta \in (0,1/2)$ to avoid pathological situations. Getting rid of the dependence on the microstate by averaging, i.e., defining
\begin{align}
	\ovl{f}_N(t) = \la f_N(t,\cdot) \ra_{\mathbb{P}_N}; && \ovl{s}_N = \la s_N \ra_{\mathbb{P}_N}, \label{KRaverage}
\end{align}
one readily calculates, most conveniently using the representation \er{alternative}, that 
\begin{align}
	\ovl{f}_N(t) - \frac{1}{2} = \left(\alpha - \frac{1}{2}\right)(1 - 2\beta)^t; && \ovl{s}_N = \beta. \label{KRaverageEvolution}
\end{align}
Hence $\ovl{f}_N(t)$ satisfies the toy Boltzmann equation 
\begin{equation}
	\ovl{f}_N(t + 1) - \ovl{f}_N(t) = \beta(1 - 2\ovl{f}_N(t)), \label{BEKR}
\end{equation}
with initial condition $\ovl{f}(0) = \alpha$ and so does 
\beq
\ovl{f}(t) := \lim_{N\raw\infty} \ovl{f}_N(t).
\eeq
This toy Boltzmann equation is actually the same as in the modified Ehrenfest model, cf.\ \er{mem_evolution}. The latter may indeed be seen as a stochastic version of the Kac ring model, in which the markers are continually randomly added and erased.
Eq.\ \er{BEKR}  is clearly irreversible, in the sense that, trivially, if $\ovl{f}_N(t)$ satisfies \er{BEKR}, then its time reversal $\ovl{f}_N(-t)$ does not (and no choice of the instantaneous time-reversal $T$ can save this in the macroscopic model). 

As in the modified Ehrenfest model, it is possible to go beyond averages and prove almost sure pointwise results, which can subsequently be converted into randomness statements. This was achieved by Hiura and Sasa (2019), Theorem 3.5. For completeness we now state and (re)prove their result, using a different argument that connects well with techniques used elsewhere in our paper. 
 Instead of \er{KRstate} the state space $A_\Z = 2^\Z\x 2^\Z$ is used, and on it the infinite analogue of \er{PNalbeta},
\begin{equation}\label{Palbeta}
	\mathbb{P}=\al^\Z\x\beta^\Z.
\end{equation}
The dynamics \er{kacevolution} remains the same, but without the periodic boundary conditions. The interpretation is now one of coloured balls moving on an infinite one-dimensional lattice $\Z$, which Hiura and Sasa (2019) call the `Kac chain'. As in the modified Ehrenfest model, the relevant macroscopic quantities are now the limits of \er{4.4} as $N\raw\infty$.
\begin{theorem}\label{HS} 
Let $\mathbb{P}\equiv \mathbb{P}_{(\al,\beta)}$ be given by \er{Palbeta}, in which
 $\alpha$ and $\beta$ are \emph{computable} real numbers. For each $t \in \N$ and each $\mathbb{P}$-random microstate $(x,y)\in A_\Z$ (in the sense of Definition \ref{defCR}), we have
\begin{align}
	\lim_{N \raw \infty} f_N(t,x,y) = f(t); && \lim_{N \raw \infty} s_N(x,y) = \beta, \label{KRmain}
\end{align}
where $f(t)$ is the solution of the toy Boltzmann equation
\begin{equation}
	f(t + 1) - f(t) = \beta(1 - 2f(t)) \label{BEKR2}
\end{equation}
with initial condition $f(0) = \alpha$.
\end{theorem}
\emph{Proof}. The arguments after \er{slln} suffice to prove the second limit in \er{KRmain} and also the first at $t = 0$, since $(x_n(0))_{n \in \Z}$ and $(y_n)_{n \in \Z}$ are both i.i.d.\ sequences by the choice of $\mathbb{P}$. For $t > 0$, this is no longer the case since correlations develop between the colours of the balls. Despite these correlations, a large deviations principle can still be proven. The key observation is that the random vector
\begin{equation}\label{history}
	(x_{n - t}, y_{n - t}, \dots, y_{n - 1})
\end{equation}
forms an irreducible Markov chain as $n$ ranges over $\Z$. By the theory of large deviations for Markov chains, the empirical measure 
\begin{equation}\label{em}
	\frac{1}{2N + 1} \sum_{n = -N}^N \delta_{(x_{n - t}, y_{n - t}, \dots, y_{n - 1})}
\end{equation}
satisfies a large deviations principle.\footnote{See Dembo and Zeitouni (1998), Section 3.1. Like in Cram\'{e}r's theorem, the corresponding rate function has a \emph{unique} zero, namely the unique stationary distribution of the chain \er{history}. This implies that the rate function of $f_N(t,\cdot)$ also has a unique zero, namely \er{zero}, which allows for the bound \er{4.18}.} Because $x_n(t)$ is a function of \er{history}, the one-particle distribution $f_N(t,\cdot)$ is a function of \er{em} and this allows us to transfer the large deviations principle of \er{em} to $f_N(t,\cdot)$.\footnote{The possibility of transferring large deviations principles is known as the \emph{contraction principle}, see Dembo and Zeitouni (1998), Section 4.2.} Setting 
\begin{equation}\label{zero}
f(t) = \la f_N(t,\cdot) \ra_{\mathbb{P}},
\end{equation}
it follows that
\begin{align}
	U_N(m)=\{(x,y)\in A_\Z: |f_N(t,x,y)-f(t)|> 1/m\}; && \mathbb{P}(U_N(m)) \leq e^{-C(m)N}. \label{4.18}
\end{align}
The remainder of the proof is then the same as in the modified Ehrenfest model:
from \er{4.18} one derives pointwise $\mathbb{P}$-almost sure convergence via the Borel--Cantelli lemma, and then pointwise convergence for each $\mathbb{P}$-random $(x,y)$ via Lemma \ref{lemma1}. Eq.\ \er{BEKR} then implies \er{BEKR2}. 
\QED
\smallskip

If one extends $\mathbb{P}$ to a probability measure on (discrete) paths in $A_\Z$, simply by taking the probability of some path to be the same as the probability of its initial condition, the situation is  analogous to the modified Ehrenfest model: each  $\mathbb{P}$-random path induces a one-particle distribution function $f(t)$ that satisfies the toy \BE\ \er{BEKR2}, whereas the time-reversed path does not (unless it is constant in time) and has exponentially small probability (and zero probability at $N=\infty$). The status of the \SZA\ in the Kac chain model is also analogous to 
the case of the 
modified Ehrenfest model, as discussed at the end of \S 3:
\begin{enumerate}
\item 
The analogue of the \SZA\ of Boltzmann (1872) is the following assumption:
  \emph{for each $t \in \N$, the proportion of white balls about to leave a marked site is equal to $\beta f(t,x,y)$ and similarly equals $\beta(1 - f(t,x,y))$ for black balls.}
 Here we have defined, cf.\ \er{KRmain},
\begin{equation}
	f(t,x,y) = \lim_{N \raw \infty} f_N(t,x,y) ,
\end{equation}
for all $t$ and $(x,y)$ for which this limit exists; by Theorem \ref{HS}, this is the case for all $\mathbb{P}$-random $(x,y)$ at all $t$. As before, if this assumption holds for $(x,y)$, then $f(t,x,y)$ satisfies the toy Boltzmann equation \er{BEKR2}. The precise formulation of the \SZA\ for white balls, then,  is
\begin{equation}
	\lim_{N \raw \infty} \frac{1}{2N + 1} \sum_{n = -N}^N x_n(t)y_n = \beta \cdot \lim_{N \raw \infty} \frac{1}{2N + 1} \sum_{n = -N}^N x_n(t),
\end{equation}
which is equivalent to
\begin{equation}
	\lim_{N \raw \infty} \frac{1}{2N + 1} \sum_{n = -N}^N x_n(t)(y_n - \beta) = 0.
\end{equation}
This  again has the form of a law of large numbers, which, using the same line of proof as \er{KRmain}, can be shown to hold for all $\mathbb{P}$-random $(x,y) \in A_{\Z}$. This shows how also in this model the \SZA\ of Boltzmann (1872) is a consequence of the deeper assumption of microscopic randomness relative to an i.i.d.\ probability distribution on microstates.
\item The necessity of i.i.d.\ initial conditions under the assumption of exchangeability follows from the general reasoning at the end of \S 2, since the Kac chain model can be formulated in terms of random variables taking values in $\{0,1\}$.

Note, however, that exchangeability is not a natural assumption for the Kac chain model, since it is immediately broken. The fundamental reason for this is that particles are labelled by their position on the lattice, rather than by arbitrary labels.
\end{enumerate}
\section{Autonomy from ergodicity}\label{EET}
One of the key features of the \BE, as well as of the various toy models studied in this paper, and more generally of emergent phenomena,  is their \emph{autonomy}: 
although these  equations are derived from microscopic data, in the limit they only depend on these data via  the macroscopic quantities they induce. The strong law of large numbers covers this phenomenon for averages of an infinite sequence of coin tosses, but here we must appeal to a more powerful and general version of this law, namely Birkhoff's famous \emph{pointwise 
ergodic theorem}, in the so-called effective (= computable) version derived in algorithmic randomness theory.\footnote{See especially Bienvenu et al.\ (2012). 
Footnote \ref{EETrefs} gives additional references. } 

Effective ergodic theory also provides at least a mathematical context for analyzing the (lack of) necessity of our (algorithmic) randomness assumptions.
If an infinite coin toss is our guide for obtaining macroscopic quantities like the average outcome $1/2$, then surely asking 
an infinite sequence to be $f^{\N}$-random (= \ML\ random) is far too strong (as simple computable examples show.\footnote{Take the sequence that repeats 01 or indeed any finite pattern averaging to 1/2 infinitely often.} 
Similarly, our previous arguments show that $\mathbb{P}$-randomness for suitable $\mathbb{P}$ is sufficient for Boltzmann-type equations; but so are $\mathbb{P}$-almost sure arguments. These are valid under weaker assumptions, because the set of all $\mathbb{P}$-random points in a computable probability space $(X, \mathbb{P})$ is just one example of a probability-one subset.\footnote{For example, take the uniform probability $f$ on $2^\N$, with induced Bernoulli probability measure $f^\N$ on $X=2^\N$. Then the set $\CR_f$ of all $f^\N$-random sequences has $f^\N(\CR_f)=1$ and its elements $s$ have the ``right'' macroscopic behaviour \er{slln}. But so does a sequence like $010101\cdots$, and one can easily construct a countable number of such computable sequences with the ``right'' macroscopic behaviour. These are of course far from $f^\N$-random.}

 As a warm-up, let us recall how the strong law of large numbers for a fair coin toss and the much stronger fact that almost every outcome sequence is Borel normal follow from Birkhoff's  pointwise ergodic theorem. This is a relevant exercise, because, as said, the former is the simplest version of what we would call (correct) macroscopic behaviour, whilst the latter provides additional pointwise almost sure macroscopic laws. The setting is a dynamical system $(X, \Sg, P,T)$ where $(X,\Sg,P)$ is a probability space and $T:X\raw X$ is a (not necessarily invertible) measurable and measure-preserving map, i.e., $P(T\inv B)=P(B)$ for each $B\in\Sg$. In addition, we assume that the dynamical system ergodic which means that $T\inv B= B$ implies $P(B) \in \{0,1\}$. Then
\begin{align}
	\lim_{N\raw\infty}\frac{1}{N} \sum_{n=0}^{N - 1}\dl_{T^nx}(B )=P(B) \label{Erg1}
\end{align}
for $P$-almost every $x\in X$ and every $B\in\Sg$, where we use the same notation as in \er{1.2}, i.e., $\dl_{T^nx}(B)=1$ if $T^nx\in B$ and $\dl_{T^nx}(B)=0$ if $T^nx\notin B$. Apply this to the case in which
\begin{align}
	X=2^\N; && \Sg=\CF; &&  P=f^\N; && T=S, \label{cointoss}
\end{align}
where $\mathcal{F}$ is defined in Appendix \ref{AR}, and $S$ is the shift map, given by $(Ss)_n=s_{n+1}$. This system is well known to be ergodic.\footnote{A good reference for all of the above is for example Viana and Oliveira (2016), Chapter 4. Mackey (1974) is a beautiful historical survey of the interaction between ergodic theory, statistical mechanics, and probability theory.} If $B_1=\{s\in 2^\N\mid s_0=1\}$, then $\dl_{S^ns}(B_1)=s_n$ and $f^\N(B_1)=1/2$, so that \er{Erg1} with $B = B_1$ reduces to the strong law of large numbers \er{slln}. More generally, take any $\sg\in 2^N$ and put $B_{\sg}=[\sg]_N$, cf.\ \er{cylinder}. Then $\dl_{S^ns}(B_{\sg}) = 1$ iff $s_{n}\cdots s_{n + N - 1}=\sg$, so that  the left-hand side of \er{Erg1} with $B=B_{\sg}$ equals the asymptotic relative frequency of the string $\sg$ in $s$. Furthermore, $f^\N(B_{\sg})=2^{-N}$, so that \er{Erg1} states that $f^\N$-almost every $s\in 2^\N$ is Borel normal.\footnote{See e.g.\ Calude (2002), Definition 6.53. Borel normality of $x\in [0,1]$ was the first serious attempt to define randomness. It was flawed because \emph{Champernowne's number} $0.12345678910111213 \cdots$  is Borel nomal (in base 10) but hardly random, in view of the simple rule behind it. Likewise, in base 2 the binary analogue of Champernowne's number, i.e.,   $0110111001011101111000\cdots\in 2^\N$, is Borel normal but computable and hence it cannot be $f^\N$-random.}

Of course, this is the kind of result we wish to sharpen according to our familiar pattern of replacing the ``for  $f^\N$-almost every $s\in 2^\N$'' clause by ``for all $f^\N$-random $s\in 2^\N$''. This can be done in several ways,\footnote{See for example Calude (2002), Theorem 6.57, for a direct argument.}
but in the spirit of the above story we invoke  the \emph{effective ergodic theorem.}\footnote{\label{EETrefs} See especially Bienvenu et al.\ (2012), Theorem 8. Similar results appear in  Galatolo,  Hoyrup, and  Rojas (2010), Theorem 3.2.2, and Franklin et al.\ (2012), Theorem 6. The first author to prove such results was V'yugin (1997). See also the reviews by Towsner (2020) and V'yugin (2022).  At the cost of additional computability assumptions, this theorem sharpens Birkhoff's pointwise ergodic theorem, which yields the first claim for $P$-almost every $x\in X$. As such, this is one of the key results in algorithmic randomness that replaces ``for $P$-almost all $x$'' by ``for all $P$-random $x$''.}
Let $(X,\Sigma,P)$ be a computable probability space (see Appendix \ref{AR}) and $T:X\raw X$ a computable (and hence continuous) measure-preserving map. Let $(X,\Sigma,P,T)$  be ergodic. Under these assumptions:\footnote{Note that 
	since $P(X)=1$ and $\dl_x(X\backslash V)=1-\dl_x(V)$,  all  assumptions as well as the statement of the theorem may equivalently be stated for computable closed sets, i.e., complements of computable opens (our sets will be clopen).}
\begin{enumerate}
	\item For each computable open set $V\subset X$ and
	all $P$-random $x\in X$ one has
	\begin{equation}
		\lim_{N\raw\infty}\frac{1}{N} \sum_{n=0}^{N - 1}\dl_{T^nx}(V)=P(V). \label{Erg} 
	\end{equation}
	\item Conversely, a point $x\in X$ is $P$-random if \er{Erg} holds for every computable open $V\subset X$.
\end{enumerate}
Since the dynamical system \er{cointoss} satisfies these assumptions, it follows from part 1 and the choice $V=[\sg]_N$ in \er{Erg} that each $f^\N$-random sequence $s\in 2^\N$ is Borel normal.  

After this preparation, we return to the Kac chain model. Like the strong law of large numbers for a fair coin toss, the macroscopic laws of Theorem \ref{HS} can be derived from the effective ergodic theorem. Our key observation is that the first equation in \er{KRmain} may be rewritten as a two-sided version of the ergodic limit \er{Erg}. To this end, take
\begin{align}
	X=A_\Z; && P=\mathbb{P}=\al^\Z\x\beta^\Z; && T = S; && V=\phv_{-t}U; && U=\{(x,y)\in A_\Z\mid x_0=1\}, \label{4.20}
\end{align} 
where $S$ is now the (double) bilateral shift $S(x,y)_n = (x_{n+1}, y_{n+1})$, $\varphi_{-t}U$ denotes the preimage of $U$ under $\varphi_t$, and $x\in X$ in \er{Erg} is now of course $(x,y)\in A_\Z$.\footnote{As is clear from our choice of dynamical system, our use of the ergodic theorem differers from traditional applications of ergodic theory to statistical physics, which apply the ergodic theorem to the microscopic time evolution, assumed to be ergodic (which is often not the case or very hard to justify).} This system is  well known to be ergodic,\footnote{And computable provided $\al$ and $\beta$ are computable reals, with $S$ easily shown to be computable.} and since $S$ is invertible (unlike its counterpart in the unilateral case), we can extend \er{Erg} to a sum from $n=-N$ to $n=N$ with normalization $1/(2N+1)$. The point is that we may now identify
\begin{align}
	\frac{1}{2N+1} \sum_{n=-N}^{N}\dl_{T^nx}(V) = f_N(t,x,y); && P(V)=\frac{1}{2} + \left(\alpha - \frac{1}{2}\right)(1 - 2\beta)^t, \label{4.21}
\end{align}
upon which \er{Erg} indeed turns into the first equation in \er{KRmain}. The reasoning for the second equation is similar, using the set $V = \{(x,y) \in A_{\Z} \mid y_0 = 1\}$. The first part of \er{4.21} follows from \er{4.4evolution} and the equalities
\begin{align} 
	S^{-n}U=\{(x,y)\in A_\Z\mid x_n=1\}; && \dl_{S^n(x,y)}(\phv_{-t}U)=\dl_{\phv_t(x,y)}(S^{-n}U) = x_n(t),
\end{align}
of which the second holds since, crucially, $S$ and $\varphi$ commute. The second  part of \er{4.21} comes down to the fact that
\begin{equation}
	\mathbb{P}(\varphi_{-t} U) = \mathbb{P}(x_0(t) = 1) = \la x_0(t) \ra_{\mathbb{P}},
\end{equation}
the average being computed using \er{alternative} in precisely the same way as in \er{KRaverage} - \er{KRaverageEvolution}. Generalizing \er{4.20} and the previous reasoning to arbitrary computable opens $U\subset A_{\Z}$, eq.\ \er{Erg} becomes
\begin{equation}
	\lim_{N\raw\infty}\frac{1}{2N+1} \sum_{n=-N}^{N}\dl_{\phv_t(x,y)}(S^{-n}U)=\mathbb{P}(\phv_{-t}U). \label{ErgK} 
\end{equation}
For each choice of $U$, the left-hand side here may be viewed as a macroscopic quantity of the Kac chain, evaluated at time $t$ under an initial microstate $(x,y)$. Under the assumption of $\mathbb{P}$-randomness, the effective ergodic theorem implies that this quantity has a time evolution given by the right-hand side. Moreover, this time evolution is \emph{autonomous}, in the sense that it does not depend on the specific $\mathbb{P}$-random microstate $(x,y)$.\footnote{However, this is not to say there is an autonomous evolution \emph{equation}, which is the case for the simplest macroscopic quantity \er{4.4evolution}, but need not be the case for more complicated ones.} By the converse of the effective ergodic theorem, these macroscopic laws characterize $\mathbb{P}$-randomness.\footnote{This characterization holds for any $t \in \N$, in particular for $t = 0$. Hence if $(x,y)$ is such that its macroscopic values are all correct at $t = 0$ in the sense of \er{ErgK}, then $(x,y)$ is $\mathbb{P}$-random, which implies that its macroscopic values will also be correct (in the same sense) at all future times $t > 0$.}

The strength of our randomness assumption can now clearly be seen. In addition to the macroscopic laws of Theorem \ref{HS}, $\mathbb{P}$-randomness implies a whole family of macroscopic laws \er{ErgK} through the effective ergodic theorem.\footnote{Bienvenu et al.\ (2012) prove two different effective ergodic theorems. Apart from \er{Erg}, which is their Theorem 8, they also show that $ \lim_{N\raw\infty}\frac{1}{N} \sum_{n=0}^{N - 1} f(T^nx)=\int_X fdP$, for arbitrary lower semicomputable functions $f:X\raw [0,\infty)$ and arbitrary $P$-random points $x\in X$ (which is their Theorem 10). These theorems are equivalent in the classical case, where $f$ is  required to be in $L^1(X,P)$, but \emph{prima facie} the left-hand side adds further macroscopic laws.} Hence, at least for the Kac chain, our notion of randomness can only be accepted in so far as one is willing to accept that many macroscopic quantities have autonomous evolutions. This raises the question of whether all these quantities are physically relevant, or whether a more restricted notion of randomness should be used in statistical physics, a question already asked by Hiura and Sasa (2019), Section 6.1.

The same analysis can be carried out for the modified Ehrenfest model. Recall that the relevant probability space is $(\Omega, \mathbb{P})$, where $\Omega = (2^{\N})^{\mathbb{T}}$ and $\mathbb{P}$ is the Markovian probability measure defined by \er{Omega}. We denote the components of $\omega \in \Omega$ by $\omega_n(t)$ and define $\omega(t) = (\omega_n(t))_{n \in \N}$. Let
\begin{align}\label{5.9}
	X = \Omega; && P = \mathbb{P}; && T = S; && V = \pi_t^{-1}U,
\end{align}
where $S$ is now the (unilateral) shift $(S\omega)_n(t) = \omega_{n + 1}(t)$, $\pi_t : \Omega \to 2^{\N}$ is the map $\omega \mapsto \omega(t)$ and $U \subset 2^{\N}$ is a computable open. Under the same assumption as Theorem \ref{t3.1}, this dynamical system is computable. It is also ergodic, with the same proof as the ergodicity of \er{cointoss}. Applying the effective ergodic theorem \er{Erg} and using as before a commutation trick, which in this case is $\pi_t \circ S = L \circ \pi_t$ with the left shift on $2^{\N}$ being denoted by $L$, results in 
\begin{equation}\label{ErgE}
	\lim_{N \to \infty} \frac{1}{N} \sum_{n = 0}^{N - 1} \delta_{\omega(t)}(L^{-n}U) = \mathbb{P}(\pi_t^{-1}U)
\end{equation}
for all $\mathbb{P}$-random $\omega \in \Omega$. Once again, we interpret the left-hand side as a macroscopic quantity and the right-hand side as its time evolution. The macroscopic law of Theorem \ref{t3.1} can be recovered by the choice $U = \{s \in 2^{\N} \mid s_0 = 1\}$.

We may now distinguish the role of the probability measure $\mathbb{P}$ itself in deriving irreversibility from the role played by our algorithmic randomness assumptions relative to $\mathbb{P}$: the former yields irreversibility of the equations associated with the ergodic theorems \er{ErgK} and \er{ErgE}, whereas the latter locates individual microscopic configurations for which these equations are satisfied. Once these configurations fall in the right class thus identified, they disappear from the equation itself; this is the whole point of pointwise ergodic theorems. Furthermore,  it is the assumption of i.i.d.\ initial conditions which makes the dynamical systems \er{4.20} and \er{5.9} ergodic in the first place, and hence, given the pointwise ergodic theorem,  implies a family of autonomous macroscopic laws. 

This  once again raises the question: what justifies i.i.d.\ initial conditions? We have already seen that under the assumption of exchangeability, i.i.d.\ initial conditions are necessary for the law of large numbers in the Ehrenfest model and the Kac chain model. Ergodic theory allows for a more general statement:
\begin{theorem}\label{iid}
	Suppose the dynamical system $(A^{\N}, \mathbb{P}, S)$, with $S$ the left shift, is ergodic and $\mathbb{P}$ is exchangeable. Then $\mathbb{P}$ is an i.i.d.\ probability measure.
\end{theorem} 
\emph{Proof}. Assuming $A$ is a sufficiently regular space,\footnote{Typically a Polish space, i.e., a separable completely metrizable topological space.}
a generalization of de Finetti's theorem states that $\mathbb{P}$ is a mixture of i.i.d.\ probability measures. It then follows from the well-known fact that ergodic measures are extremal in the space of invariant measures that $\mathbb{P}$ must itself be i.i.d.\footnote{See Aldous (1985), Section 12 for more details. For the generalization of de Finetti's theorem, see Hewitt and Savage (1955), in which it is proven for compact Hausdorff spaces, from which the Polish case follows.} \QED
\smallskip

How does this generalize our reasoning at the end of \S\ref{EM}? Ergodicity is actually equivalent to the ergodic theorem in the form \er{Erg1},\footnote{Suppose \er{Erg1} holds and $B \in \Sigma$ satisfies $T^{-1}B = B$. It follows that $x \in B$ iff $T^n x \in B$ for all $n \geq 0$. Hence the sum on the left-hand side of \er{Erg1} is either 0 or 1 and the same must hold for the right-hand side.} hence Theorem \ref{iid} states that, assuming exchangeability for epistemic reasons and assuming that autonomy of the macroscopic equation requires the ergodic theorem in the way we used it (as opposed to e.g.\ Boltzmann's completely different use of ergodicity in the embryonic version he coined, namely in his attempts to justify the microcanonical ensemble),\footnote{See  Darrigol (2018) and Uffink (2007, 2022). It is hard to pinpoint to a single  paper by Boltzmann on this point.}
i.i.d.\ initial conditions are necessary. However, while mathematically necessary, i.i.d.\ initial conditions would only be physically justified if it can be shown that any (reasonable) initial condition converges to an approximately i.i.d.\ probability distribution.\footnote{This ``generation of chaos" approach is taken up by Lukkarinen and Vuoksenmaa (2024) for a stochastic model.} This seems a promising approach to the justification of such initial conditions, but it remains to be shown that their stochastic mechanism for the  generation of chaos also holds for deterministic models.\footnote{Here the ``right'' initial conditions are crucial, think of velocity inversion \`{a} lo Loschmidt.}  In any case, it would be welcome if it could be shown  that Boltzmann-like equations also hold if the  initial probability distribution is only approximately i.i.d.

Finally, how widely applicable is our use of the effective ergodic theorem?  The crucial assumption in the derivation of \er{ErgK} and \er{ErgE} is commutativity of the dynamics and the shift $S$ on the state space, that is, $\varphi_t \circ S = S \circ \varphi_t$ in the Kac chain model and $\pi_t \circ S = L \circ \pi_t$ in the Ehrenfest model. However, this assumption is quite strong, since it implies that the state of the system is ergodic at all times $t > 0$ if it is i.i.d.\ at $t = 0$.\footnote{In the deterministic case, let $\mathbb{P}_t = \mathbb{P} \circ \varphi_{-t}$ be the probability measure at time $t$. If $(A^{\N}, \mathbb{P}, S)$ is ergodic and $\varphi \circ S = S \circ \varphi$, then $(A^{\N}, \mathbb{P}_t, S)$ is also ergodic. To see this, suppose $S^{-1}B = B$ for some measurable $B \subset A^{\N}$. Applying $\varphi_{-t}$ and making use of commutativity, we get $S^{-1}\varphi_{-t} B = \varphi_{-t} B$ and hence $\mathbb{P}_t(B) = \mathbb{P}(\varphi_{-t}B) \in \{0,1\}$.} For models that preserve exchangeability this would imply by Theorem \ref{iid} that the state at any time $t > 0$ is i.i.d. We conclude that the results of this section can only hold in two cases:
\begin{enumerate}
	\item Models in which all particles are independent, typified in our paper by the modified Ehrenfest urn model.  Commutativity then represents the fact that the dynamics of the remaining particles does not change when neglecting some particles.
	\item Models in which exchangeability is not preserved,  typified in our paper by the Kac chain model (and more generally by cellular automata). In such models commutativity of dynamics and shift represents spatial homogeneity of the model.
\end{enumerate}
This excludes most realistic models of kinetic theory, notably hard-sphere dynamics, since these preserve exchangeability \emph{and} have the property that correlations develop between particles. Nonetheless, seeking the mathematical origin of the autonomy of macroscopic evolution equations (even short of their irreversibility) in pointwise ergodic theorems seems a very attractive scenario, which might also be realized in  different ways that are not limited by the above restrictions. 
\section{Conclusion}
Summarizing the lessons from our three cases, the ensuing picture of the irreversibility of macroscopic Boltzmann-like equations from reversible microscopic dynamics  
is as follows:
\begin{enumerate}
	\item In all rigorous derivations, functions $f(t,\cdot)$ supposed to solve a Boltzmann-like equation start their lives as  sequences $(f_N)$ of random variables on a space of $\sim N$ microscopic degrees of freedom (d.o.f.) equipped with a probability measure $\mathbb{P}_N$. Most derivations of the full \BE, including those by
Lanford (1975) for short times and recently Deng, Hani, and Ma (2024) for arbitrary times, define $f(t,\cdot)$
as a suitable (Boltzmann--Grad) limit $N\raw\infty$ of the \emph{average} $\la f_N\ra_{\mathbb{P}_N}$, which coincides with the first (single-particle) correlation  function in the BBGKY hierarchy.

Following Hiura and Sasa (2019) we propose to replace these finite-$N$ \emph{averages} over the microscopic d.o.f.\ by a \emph{pointwise} limit with respect to a single (individual) microscopic configuration $x$ of some suitable infinite-volume (i.e., $N=\infty$) idealization of the microscopic model,  and, in the spirit of the strong law of large numbers or the more general pointwise ergodic theorem, wish to prove that this pointwise limit exists and is independent of $x$ whenever $x$ is algorithmically random with respect to the infinite-volume limit $\mathbb{P}$ of $\mathbb{P}_N$ (if this limit measure exists). We were able to carry out this program for at least our two toy models, i.e., the (modified) stochastic Ehrenfest model and the deterministic Kac ring model, and will hopefully be able to so in the future also for real dilute gases. Bracketing the incompleteness theorems for algorithmic randomness for the moment (see below), this picture is conceptually quite satisfying, especially in giving an explanation of the autonomy of the ensuing macroscopic evolution equations.
	\item The choice of $\mathbb{P}_N$ under which  Lanford and his followers  derive the \BE\ makes the positions and velocities of the $N$ particles comprising the gas i.i.d.\ random variables at some initial time $t=0$. As explained in \S\ref{OBE}, this is one of the  sources of the irreversibility of the \BE--the others were \emph{coarse-graining}, \emph{limits}, and \emph{specific dynamics}.  These sources act in concert:  one cannot single out one of them as \emph{the} reason for irreversibility. That said, the i.i.d.\ assumption at $t=0$, in either the ``ensemble'' form based on averages used in most literature or in our preferred ``individual'' form based on (effective) pointwise ergodic theorems and algorithmic randomness, seems the hardest one to justify.\footnote{The most detailed  attempt to do so we are aware of is Cercignani, Illner, \& Pulvirenti (2013), p.\ 25. Their first justification is
 that `in general, we cannot handle the single particles but rather act on the gas
as a whole at a macroscopic level, usually starting from an equilibrium state (for which the i.i.d.\ property holds).'  But for all interesting initial states the gas is out of equilibrium! The second is that  `if
we choose the initial data for the particles at random there is an overwhelming probability that [the i.i.d.\ property] is satisfied for $t = 0$.' This seems to refer to some  probability measure over the space of probability measures over initial conditions that is maximized by i.i.d.\ conditions, but we find no details and in any case such an approach is remote from our own attempts to look at individual configurations.}
	
One justification is simply an appeal to ``succes'': making this assumption, in concert with the other ideas just mentioned, implies the \BE\ and hence macroscopic irreversibility at least with respect to the one-particle distribution function $f(t,\cdot)$. As just mentioned, this is the case for the averaged-out distribution function appearing in the full \BE, and holds pointwise (and hence individually) for our toy models.\footnote{ Spohn (1991), p.\ 49,  already pointed out that the i.i.d.\ initial conditions `presumably are too strong and it is of interest to weaken them. But these conditions suffice the derive the \BE.'}

Another is that, as discussed for the modified Ehrenfest model at the end of \S\ref{EM} and for the Kac ring model at the end of \S\ref{KRM}, under the (at least epistemically justified) assumption of exchangeability, one is practically  forced into the i.i.d.\ assumption at $t=0$ by the goal of an autonomous macroscopic equation (although the reasons for this are somewhat different in the two models). 
We  hope that something similar is also the case for dilute gases.

Whatever their origin, it is tempting to see the apparently crucial assumption of i.i.d.\ assumptions as some version or implementation of the \emph{past hypothesis}, according to which, very briefly, the universe was in a low entropy state in the remote past.\footnote{See Albert (2000),  Earman (2006), and Callender (2024) and references therein.} Indeed, there is some overlap in the emphasis that both approaches put on special initial conditions as one of the sources of irreversibility. On the other hand, there are at least two major differences. First, the i.i.d.\ initial conditions for the \BE\ and related toy models are not cosmological in origin; they describe gases in small containers prepared by engineers; and even if these initial conditions are applied to something like a cosmological fluid, the real low entropy assumption would lie in the choice of the initial one-particle distribution function $p(\mathbf{z})$ in \er{1.4}, rather than in the product structure of \er{1.4}. Second, the past hypothesis is often proposed as \emph{the} source of irreversibility (or even the arrow of time in case these are conflated), whereas we  see the i.i.d.\ initial conditions as merely \emph{one} of these sources.
\item
Finally, although it is hard to pin this down more precisely (and more historical research is needed to do so), throughout the history of (kinetic) gas theory from  Jan Baptista van Helmont (1579--1644) onwards, who introduced the word `gas' by a direct appeal to the Greek concept of `chaos',\footnote{See  \url{https://nl.wikipedia.org/wiki/Jan_Baptista_van_Helmont}. On page 59 of his posthumous work \emph{Ortus medicinae, id est Initia physicae inaudita} from 1652 he writes: '\emph{in nominis egestate, halitum illum, Gas vocavi, non longe a Chao}', meaning: `since I needed a name for this vapour I called it ``gas'', resembling ``chaos''.'} to Maxwell and Boltzmann, whose German word \SZA\ was soon translated into English as the hypothesis of `molecular chaos',\footnote{As noted by Uffink and Valente (2015), there is some unfortunate confusion in the literature between 
the \SZA, which only asks for factorization of particles about to collide, and the much stronger
i.i.d.\ assumption, both of which are referred to as (the hypothesis of) molecular chaos. At this point we discuss the latter.
See also  Pulvirenti \& Simonella (2016) for some useful historical comments on the concept of chaos in kinetic theory.}
 there has been a strong intuition that particles in a gas move ``randomly'' or ``chaotically''. And this is exactly what we try to capture in proposing that the initial configuration is $\mathbb{P}$-random with respect to some i.i.d.\ probability measure $\mathbb{P}$ on the space of microscopic degrees of freedom.
 
 However, developing this intuition into a sound mathematical argument one is walking a tightrope, since on the one hand such random motion (and hence the initial condition?)  is supposed to come from collisions, while on the other hand collisions create correlations and hence  destroy the $t=0$  i.i.d.\ property at $t>0$. This dilemma is well expressed as follows:
 \begin{quote}
 \begin{small}
 The molecular chaos  is clearly a property
of randomness. Intuitively, one feels that collisions exert a randomizing
influence, but it would be completely wrong to argue that  statistical
independence is a consequence of the dynamics.
It is evident that the chaos property,  if initially present, is almost
immediately destroyed if we insist that it should be valid everywhere. 
(Cercignani, Illner, \& Pulvirenti, 2013, p.\ 24) 
\end{small} 
\end{quote}
\end{enumerate}

Another worry lies in the role of the idealization $N=\infty$ we rely on in order to present precise results using algorithmic randomness. This concern is justified because the property of $P$-randomness we  use for infinite computable spaces like $X=2^\N$ (see Appendix \ref{AR})  by imposing it on microscopic states $x\in X$, collapses if $X$ is finite, since in that case
each  point $x\in X$ is $P$-random.\footnote{See Hertling and Weihrauch (2003), Example 3.6.1.
Here we assume for simplicity that  $P(x)>0$ for all $x\in X$.}
 Thus one faces the danger of violating what is sometimes called
\emph{Earman's Principle}:
 \begin{quote}
 \begin{small}
`While idealizations are useful and, perhaps, even essential to progress in physics, a sound principle of interpretation would seem to be that no effect  can be counted as  a genuine physical effect if it disappears
when the idealizations are removed.'  (Earman, 2004, p.\ 191)
\end{small}
\end{quote}
This often applies to mathematical physics, where infinite idealizations are liked because they typically give crystal-clear mathematical results.\footnote{There is of course a huge literature on  infinite idealizations. Palacios \& Valente (2021) is a good starting point.}  
Likewise, one has a \emph{Paradox of Infinite Limits}:\footnote{Although we are not concerned with realism, this `Paradox' poses a similar threat to empiricists. See Landsman (2026a) for the claim that realism is incompatible with at least the history of mathematical physics.}
 \begin{quote}
 \begin{small}
 On the one hand, a scientific
realist must believe that real physical systems are finite, as it is indeed suggested by some of our most
successful background theories such as the atomic theory of matter and general relativity. On the other
hand, she must believe that the content of scientific theories invoking infinite idealizations is true or
approximatively true insofar as they are indispensable to recover empirically correct results.
(Palacios \& Valente, 2021, Introduction)
\end{small}
\end{quote}
This applies, for example, to phase transitions and spontaneous symmetry breaking, which  in nature  clearly occur in finite systems but which official mathematical physics only allows in infinite systems. The answer from sound mathematical physics is that one should take this challenge seriously and invent new arguments that justify the use of idealizations by showing what happens in the corresponding finite systems. This strategy is expressed well by 
 \emph{Butterfield's Principle}:\footnote{This terminology seems to have been introduced in Landsman (2017), page 12. }
\begin{quote} \begin{small}
``there is a weaker,
yet still vivid, novel and robust behaviour that
occurs before we get to the limit, i.e. for finite $N$. And it is this weaker behaviour
which is physically real.'' (Butterfield, 2011, p.\ 1065)
\end{small}\end{quote}
For example, spontaneous symmetry breaking is driven in finite systems by a very subtle mechanism involving a combination of Anderson's `tower of states' and an increased sensitivity to perturbations as the system grows large.\footnote{See Landsman (2017), Tasaki (2020), and van de Ven et al.\ (2020).} This has yet to be taken up in the philosophy of physics. 

 In the same spirit, in our context of algorithmic randomness one may remove the idealization by invoking a genuine new idea compared to $P$-randomness, namely \emph{Kolmogorov randomness} for finite $N$, which in fact historically predated the concept \ML\ randomness of which $P$-randomness is a direct generalization (see Appendix \ref{AR}). The continuity between Kolmogorov randomness and \ML\ randomness is expressed by Theorem \ref{LG}, but admittedly much remains to be done in actually implementing Butterfield's Principle on that basis. In particular, for finite $N$ one will not obtain the exact \BE, neither for dilute gases nor in our toy models, but the approximate equation one does obtain should still be irreversible in a similarly approximate sense, namely that this time the time-reversed solutions (in the sense of the \emph{first} bullet near the end of \S\ref{OBE}) are not impossible but just unlikely. 

We close with two comments on  algorithmic randomness in connection with irreversibility.

First, no assumption of algorithmic randomness stands alone; as our term `$P$-randomness' suggests, this is always defined relative to some probability measure $P$, whose specific form is an assumption. This form is what, in conjunction with the other sources discussed above,  may (or may not) lead to (macroscopic) irreversibility and possible connotations with chaos.  The further identification of $P$-random initial conditions or trajectories then allows one to pass to an individual description, in making it possible to say that \emph{this} microscopic initial condition or trajectory is random (and then, for suitable $P$, gives rise to irreversible macroscopic evolution). In addition, it enables statements of the kind made for example after \er{reversed}, to the effect that if some microscopic path is $P$-random, then the corresponding time-reversed path is not (and hence is ``disabled'').

Second,  how \emph{explicit} is our identification of microstates giving rise to irreversible macroscopic behavious as being ``algorithmically random''? This is a fascinating question, with two sides:
\begin{itemize}
	\item  On the one hand, the property of $P$-randomness of points in a computable probability space $(X,P)$ is clearly and distinctly defined (see Appendix \ref{AR}), and some individual $P$-random objects can similarly  be defined by some logical description (in standard set theory).\footnote{The most famous example is Chaitin's   $\Omega$, defined as the halting probability $\Omega=\sum_{p\mid p\ \text{halts}} 2^{-|p|}$, where the sum is over all self-delimiting  programs $p$ that can run on a given Turing machine, cf.\ Downey and Hirschfeldt (2010), \S3.13. }
	\item  On the other hand, \emph{Chaitin's second incompleteness theorem} states that if $s\in 2^{\N}$ is $f^{\N}$-random, 
	then the value of all but finitely many digits is undecidable (\`{a} la G\"{o}del) within any sufficiently comprehensive mathematical theory like ZF or even ZFC.\footnote{See Calude (2002), Theorem 8.7. This is stated for Chaitin's  $\Omega$, but as pointed out by Landsman (2021), and later acknowledged by Calude and other experts,  the proof holds for any  $f^{\N}$-random sequence. }
	\end{itemize}
Thus although ``most'' infinite binary sequences are random with respect to the probability measure $f^{\N}$,  none of these can be explicitly \emph{known} or \emph{shown};  for doing so would blast their randomness (likewise, a spy who identifies himself is no longer a spy). 
Returning to the previous discussion on idealization, this is not an artifact of the idealization inherent in the use of \emph{infinite} binary sequences: for finite binary strings,
	\emph{Chaitin's (first) incompleteness theorem} states that although countably many strings  \emph{are} random, this can be \emph{proved} only for finitely many of these.\footnote{Raatikainen (1998) contains a very clear discussion of this theorem. } This elusiveness of algorithmically random objects (which, based on undecidability,  goes far beyond their obvious non-computability) seems  a deep and unavoidable aspect of their existence. 
\appendix
 \section{Algorithmic randomness}\label{AR}
 Triggered by ideas of Kolmogorov, the concept of $P$-randomness (where $P$ is a probability measure) was introduced by \ML\ (1966) for the  case $X=2^{\N}$ of infinite binary sequences with  the uniform Bernoulli measure $P=f^\N$. 
 Before we formally state a more general definition  of $P$-randomness, let us give the main idea (as we see it).
  In  the simplest case of \emph{statistical hypothesis testing} one tries to decide between two hypotheses, which in this case are the null hypothesis $\mathsf{H}_0$ stating that some point $x\in X$ is \emph{not} random \emph{relative to some probability measure $P$ on a set $X$},\footnote{All attempts to define some absolute concept of randomness having failed; see van Lambalgen  (1987).  Porter (2012), Eagle (2021), and Landsman (2020) provide further history, philosophy,  and analysis. } and the alternative hypothesis $\mathsf{H}_1$ that it is. Such a hypothesis is tested by specifying some subset $V\subset X$, literally called a \emph{test} in the literature, such that $\mathsf{H}_0$ is accepted if $x\in V$ and rejected if $x\notin V$. 
  
 The idea is that $V$ contains points in which some pattern occurs that compromises the randomness of these points (relative to $P$). For example, for $X=2^\N$ with $P=f^\N$ this pattern could be an unusually large number of zeros that would make its asymptotic relative frequency different from the $\half$  that is to be expected on the basis of $f^\N$. To make this precise, one should index the test $V$ by  $n\in\N$, so that, in this example, the first $n$ digits of $s\in 2^\N$ are zero. This gives $P(V_n)=2^{-n}$, and the correct way of stating that $s$ is random at least in the very narrow sense of having no unduly large fraction of zeros, is $s\notin \bigcap_n V_n$. Of course, this is not nearly enough for randomness: all other patterns have to be tested for, too, where part of the problem of defining randomness is making precise what is even meant by a ``pattern''. The latter is done via computability theory as detailed below, and the outcome is that a randomness test is now defined as a suitably computable sequence $(V_n)_n$ of open subsets $V_n\subset X$ (assuming that $X$ is a `computable topological space' and $P$ is also computable) such that $P(V_n)\leq 2^{-n}$. Some $x\in X$  passes \emph{that} test if $x\notin \bigcap_n V_n$, and $x$ is  declared to be $P$-random, or random relative to $P$, if $x\notin \bigcap_n V_n$ for \emph{all} randomness tests  $(V_n)_n$.
 
 That said, our Definition \ref{defCR} involves a slight technical reformulation of this idea, in which the above \emph{\ML\ tests} are replaced by \emph{Solovay tests}. This is purely a matter of convenience, as the latter are much simpler to use in practice, at least in our context.\footnote{See for example Theorem 6.37 in Calude (2002) or Theorem 6.2.8 in Downey and Hirschfeldt (2010)
 for the equivalence between randomness defined via \ML\ tests and  randomness defined via Solovay tests.} Furthermore, in both of these versions, specializing Definition \ref{defCR} below to $(2^\N,f^\N)$ seems to lose the above intuition based on initial segments;  but unfolding the computability requirement would reinstore that intuition.\footnote{See Li and Vit\'{a}nyi (2008), \S2.5.2, for further information. Calude (2002), \S 6.2, shows that for $X=2^\N$ (and more generally for $A^\N$ with finite $A$) one may assume that $V_N$ is a cylinder set, i.e., it is essentially defined in $2^*$.} Finally, apart from our use Solovay tests the following definition is due to  Hertling and Weihrauch (2003), \S 3, which reference also explains the `computability' of $P$ (whose definition we omit).\footnote{
There is an alternative formalism of computable metric spaces due to Hoyrup and Rojas (2009), which seems equivalent at least in the cases like $X=2^\N$ etc.\ studied in this paper. }
 \begin{definition}\label{defCR}
\begin{enumerate}
\item A \emph{computable} (or \emph{effective}) topological space $X$ is a topological space with a countable base $\mathcal{B}\subset\CO(X)$, numbered by a bijection $B:\N\stackrel{\cong}{\raw} \mathcal{B}$. In such a space:
\begin{enumerate}
\item  An open set $V\in\CO(X)$ in $X$  is \emph{computable} (effective) if,  for some (total) computable function $f:\N\raw\N$,
\beq
V=\bigcup_{n\in\N} B(f(n)).
\eeq
 \item  A \emph{sequence} $(V_n)$ of open sets $V_n\in\CO(X)$ is \emph{uniformly computable} if,  for some  (total) computable function $g:\N\x\N\raw\N$,
\beq
V_n=\bigcup_{m\in\N} B(g(n,m)).
\eeq
\end{enumerate}
\item A \emph{computable  probability space}
$(X,P)$ is a computable topological space $X$  with a computable probability measure $P$ (we suppress the bijection $B$ in the notation).
\item A (Solovay) \emph{test} in such a space is a uniformly computable sequence of opens $(V_n)$ for which 
\beq
\sum_n P(V_n)<\infty.\label{Solovay}
\eeq
\item \label{Sr} A point $x\in X$ is \emph{$P$-random} if for each test $(V_n)$ one has $x\in V_n$ for only finitely many $N$.
 \end{enumerate}
\end{definition} 
As before, a key example is $X=2^{\N}$  with $2=\{0,1\}$. Its elements are maps $s:\N\raw\{0,1\}$, and the ``right'' topology $\CO(2^\N)$ on this set (for our purposes) is generated by the so-called \emph{cylinder sets}
 \beq
 [\sg]_N=\{s\in 2^{\N}\mid s_{|N}=\sg\}, \label{cylinder}
 \eeq
   where $N\in\N$ and $\sg\in 2^N$,  and  $s_{|N}\in 2^N$ is the restriction of $s$ to $\{0, \ldots, N - 1\}$. 
 These cylinder sets, which by definition form a base $\mathcal{B}$ of the topology $\CO(2^\N)$, form a countable subset of the power set of $2^\N$ which may be numbered 
  for example via the  lexicographical order  $L:\N\stackrel{\cong}{\raw} 2^*\equiv\bigcup_{N\in\N}2^N$,
i.e., the set of all finite binary strings. This numbering, in turn, gives 
 a bijection 
\begin{align}
B:\N\stackrel{\cong}{\raw} \mathcal{B}; && n\mapsto L(n)2^{\N},\label{BL}
\end{align}
which may be used to turn $2^\N$ into a computable space. In the usual approach to probability theory on $2^\N$ the cylinder sets likewise generate the standard $\sg$-algebra $\CF$ on which probability measures may be defined; 
 $\mathcal{F}$ is also the smallest $\sg$-algebra that makes all  evaluation maps 
 \begin{align}
 X_n: 2^\N\raw \{0,1\}; &&  X_n(s)=s_n,
 \end{align}
  measurable, where $s\in 2^{\N}$ and $n\in\N$. By the Carath\'{e}odory extension theorem, any probability measure on $\CF$ is determined by its values on  $\mathcal{B}$. For example, a prior $q$ on $\{0,1\}$ (specified by $q(1)\in [0,1]$ so that $q(0)=1-q(1)$), such as the 
 uniform (unbiased) probability $q=f$, i.e., 
 \beq
 f(0)=f(1)=1/2, \label{deff}
 \eeq
  induces the corresponding Bernoulli probability measure  $q^{\N}$ on $2^{\N}$ via
  \begin{align}
  q^{\N}([\sg]_N)=q^N(\sg)=\prod_{n=0}^{N - 1} q(\sg_n). \label{defqN}
 \end{align}
In particular, $f^\N([\sg]_N)=2^{-N}$ for any $\sg\in 2^N$. The probability space $(2^\N,q^\N)$ with countable base numbered by \er{BL} is computable iff $q$ is computable (as a real).  This is all we need.

We now relate the above  definition of $P$-randomness to Kolmogorov's original concept of randomness of \emph{finite} binary strings $\sg$ was quite intuitive,\footnote{The original references are Kolmogorov (1965, 1968). Li and Vit\'{a}nyi (2008) is the standard textbook in the field.} namely that $\sg$  is random iff it is incompressible, i.e., $\sg$ is random if its shortest description is $\sg$ itself. Here the vague concept of a ``description'' is once again made precise using computability theory; roughly speaking, $\sg$ is random if the shortest computer program that outputs $\sg$ is at least as long as $\sg$ itself. More technically, 
the \emph{prefix Kolmogorov complexity} $K_T(\sg)$ of $\sg\in A^*$ is defined as the length of the shortest  program running on some universal prefix Turing machine $T$ that outputs $\sg$ and then halts:\footnote{That is, the domain $D(T)$ of $T$
consists of a prefix subset of $2^*$, so if $x\in D(T)$ then $y\notin D(T)$ whenever $x\prec y$.} fix some  universal prefix Turing machine $T$, and
 define 
 \beq
 K_T(\sg):=\min\{|x|: x\in 2^*, T(x)=\sg\},
 \eeq i.e., the length of the shortest program running on $T$ that outputs $\sg$. Different choices of $T$ give different numbers $K_T(\sg)$ only up to a $\sg$-independent constant; reflecting this arbitrariness we call
   $\sg\in A^*$  \emph{$c$-Kolmogorov random} (relative to $T$), where $c\in\N$ is some  $\sg$-independent constant,  iff
   \beq
   K_T(\sg)\geq |\sg|-c.
   \eeq
    On the other  hand, being defined as a minimum, for any string $\sg$ the number $K_T(\sg)$  can't be much longer than $|\sg|$, so that roughly speaking, for long strings (such that $c$ is insignificant relative to $|\sg|$)  $\sg$ is (Kolmogorov) random iff $K(\sg)\approx |\sg|$, omitting the reference to $T$. 
One justification of Definition \ref{defCR}.\ref{Sr}, then, is a theorem due to Schnorr stating that an infinite binary sequence $s\in 2^\N$ is  
   $f^\N$-random iff each finite initial segment $s_{|N}\in 2^N$ is uniformly Kolmogorov random,\footnote{See Downey and Hirschfeldt (2010), Theorems 6.2.3 and 6.2.8.}
    in the sense that 
   there is a constant $c\in\N$ (which may depend on $s$ but not on $N$)
such that for all $N$,
\begin{equation}
K(s_{|N})\geq N-c.  \label{EP2}
\end{equation}
  There is also a version of this result for  for arbitrary computable measures $P$ on $2^{\N}$, viz.
\begin{theorem}[Levin, G\'{a}cs]\label{LG}
Let $P$ be a computable probability measure on $2^{\N}$. Then $s\in 2^{\N}$ is $P$-random iff  there is a constant $c\in\N$ such that
  for all $N$,
\begin{equation}
K(s_{|N})\geq -\log_2( P([s_{|N}]))-c.  \label{EPLG}
\end{equation}
\end{theorem} 
If $P=f^{\N}$, then $P([s_{|N}])=2^{-N}$,  and we recover Schnorr's theorem just quoted. See  G\'{a}cs (1979). 
\section{Time reversal}\label{AT}
 In deterministic theories defined over $\R$ as the time axis, with state space $\CS$, time evolution is given by an $\R$-action on $\CS$, providing trajectories $x\mapsto x(t)\equiv \phv_t(x)$. Following Roberts (2022), we
 say that this dynamics is \emph{time  symmetric} (or \emph{time-reversal invariant}) if the following is true:
 \begin{itemize}
\item  There is  an  invertible \emph{instantaneous time-reversal map} $T:\CS\raw\CS$ with $T\inv=T$, such that 
 \begin{align}
\phv_{t}\circ T=T\circ\phv_{-t}. && (t\in\R).\label{defT}
\end{align}
\end{itemize}
 In that case, define the time-reversed path $x^R(t)$ of $x(t)$, running from   $x_0=x(0)$ to $x_f=x(\tau)$, by
\beq
x^R(t):=T\phv_{-t}(x_f), \label{3.15}
\eeq
which runs from 
$x^R(0)\equiv x^R_0=Tx_f$ to  $x^R(\tau)=Tx_0$. The reason for time  symmetry, then, is that because of \er{defT}
the path $x^R(t)$ is also given by $x^R(t)=\phv_t(x^R_0)$, so that, being given by $\phv_t$,  it 
satisfies the equations of motion, too (which are supposed to be implicit in specifying the dynamics $\phv_t$). All of this also works if $t\geq 0$ and even if $t\in\N$, in which case we constrain $t$ to $t\in [0, \tau]$ and use $\tau - t$ instead of $-t$ in \er{3.15}.

For example, in Newtonian physics
one has $x=(q,v)$ and $T$ is the  velocity inversion map
\beq
T(q,v)=(q,-v).
\eeq
 In quantum mechanics on $\R^d$ one has $\psi\in L^2(\R^d)$ and $T$ is famously complex conjugation, i.e., 
 \beq
 T\psi=\ovl{\psi}.
 \eeq 
In the non-deterministic case we restrict ourselves to time-homogenous Markov chains $(X(t))_{t\in\N}$ with state space $\CS$ (and  time $t\in\N$). First, note that   the concept of a Markov chain (or  process) is by itself independent of the direction of time:\footnote{See e.g.\ Br\'{e}maud (2020), Remark 2.1.5.} this is because the defining Markov condition 
\beq
\mathbb{P}(X(t)=x\mid X(t_1)=x_1, \ldots, X(t_n)=x_n)=\mathbb{P}(X(t)=x\mid X(t_1)=x_1),
\eeq for all $n$, $i=1, \ldots, n$, all $x_i$,  and $t_i$ such that 
$t>t_1>\cdots > t_n$, is equivalent to the same condition  for $t<t_1<\cdots < t_n$.
Trajectories, then,  are sample paths $(X(0)=x_0, \ldots,  X(\tau)=x_f)$, whose time  reversal according to \er{3.15} is $(X(0)=Tx_f, \ldots, X(\tau)=Tx_0)$. The standard definition of time  symmetry of a Markov chain is to ask that, for all times $(t_0, \ldots, \tau)$ and states $(x_0, \ldots, x_f)$,
\begin{equation}
\mathbb{P}(X(0)=Tx_f, \ldots, X(\tau)=Tx_0)=\mathbb{P}(X(0)=x_0, \ldots, X(\tau)=x_f).\label{3.16}
\end{equation}
This is easily shown to be equivalent to  the chain satisfying a twisted detailed balance condition
\begin{equation}
\mu_{Ty} P_{TyTx}= \mu_xP_{xy},\label{3.17}
\end{equation}
where $P_{xy}=\mathbb{P}(X(1)=y\mid X(0)=x)$ and $\mu_x=\mathbb{P}(X(0)=x)$ as usual. If $T=\mathrm{id}$, this simply reads
\beq
\mu_y P_{yx}=\mu_xP_{xy}. \label{3.17b}
\eeq 
Summing over $x$ gives $\mu=\mu P$ (since $P$ is a stochastic matrix), so that $\mu$ is a stationary distribution.  Hence  \er{3.16} forces the  distribution $\mu(t)$ of $X(t)$ to be  stationary and hence equal to the  distribution $\mu\equiv\mu(0)$ of $X(0)$. For example, eq.\ \er{3.17b} holds for both the microscopic and the macroscopic Ehrenfest models, where $T=\mathrm{id}$, and indeed each has a
 unique stationary distribution given by \er{pa2N} and \er{binEF}, respectively.\footnote{In both cases the stationary distribution is unique because the models are irreducible as Markov chains.}  Hence condition \er{3.16} is satisfied in both versions of the model.  
 \subsubsection*{Acknowledgements}
  The first author was funded by the NWA ORC programme
``Emergence at All Scales''. The authors are indebted to Mark Peletier, Herbert Spohn, Jos Uffink, and
two anonymous referees for very useful feedback. 
\addcontentsline{toc}{section}{References}
\begin{small}

\end{small}
\end{document}